\documentclass[prd,amsmath,amssymb,preprintnumbers,superscriptaddress,twocolumn,10pt,nofootinbib,showkeys,showpacs]{revtex4-1} %,showkeys,showpacs
\pdfoutput=1
\usepackage{graphicx}
\usepackage{subfigure}
\usepackage{float}
\usepackage{longtable}
\usepackage{lineno}
\usepackage{dcolumn}
\usepackage{bm}
\usepackage{amssymb}
\usepackage{latexsym}
\usepackage{booktabs}
\usepackage{amsmath}
\usepackage{multirow}
\usepackage{url}
\usepackage{changes}
\usepackage{soul}
\usepackage{color, xcolor}
\usepackage[colorlinks=true, linkcolor=red, citecolor=blue]{hyperref}
\usepackage{orcidlink}
\usepackage{etoolbox}

\usepackage[normalem]{ulem}
\usepackage{array}
\usepackage{enumerate}

\def\be{\begin{equation}}
\def\ee{\end{equation}}

\def\bea{\begin{eqnarray}}
\def\eea{\end{eqnarray}}

\newcommand{\apjl}{Astrophys. J. Lett.}
\newcommand{\apjs}{Astrophys. J. Suppl. Ser.}
\newcommand{\aj}{Astron. J.}
\newcommand{\mnras}{Mon. Not. R. Astron. Soc.}
\newcommand{\aap}{Astron. Astrophys.}

\newcommand{\pasp}{Publ. Astron. Soc. Pac.}
\newcommand{\aapr}{Astron. Astrophys. Rev.}
\newcommand{\pasa}{Publ. Astron. Soc. Aust.}
\newcommand{\physrep}{Phys. Rep.}
\newcommand{\pasj}{Publ. Astron. Soc. Jpn.}

\begin{document}
%\linenumbers
\title{Unveiling the spectral morphological division of fast radio bursts with CHIME/FRB Catalog 2}

\author{Wan-Peng Sun\orcidlink{0000-0001-5950-7170}}
\affiliation{Liaoning Key Laboratory of Cosmology and Astrophysics, College of Sciences, Northeastern University, Shenyang 110819, China}

\author{Yin-Long Cao\orcidlink{0009-0002-7788-8821}}
\affiliation{Liaoning Key Laboratory of Cosmology and Astrophysics, College of Sciences, Northeastern University, Shenyang 110819, China}

\author{Yong-Kun Zhang\orcidlink{0000-0002-8744-3546}}
\affiliation{National Astronomical Observatories, Chinese Academy of Sciences, Beijing 100101, China}

\author{Ji-Guo Zhang\orcidlink{0009-0004-5607-9181}}
\affiliation{Liaoning Key Laboratory of Cosmology and Astrophysics, College of Sciences, Northeastern University, Shenyang 110819, China}

\author{Xiaohui Liu\orcidlink{0000-0002-2552-7277}}
\affiliation{National Astronomical Observatories, Chinese Academy of Sciences, Beijing 100101, China}
\affiliation{School of Astronomy and Space Science, University of Chinese Academy of Sciences, Beijing 100049, China}

\author{Yichao Li\orcidlink{0000-0003-1962-2013}}\thanks{Corresponding author: liyichao@neu.edu.cn}
\affiliation{Liaoning Key Laboratory of Cosmology and Astrophysics, College of Sciences, Northeastern University, Shenyang 110819, China}

\author{Fu-Wen Zhang\orcidlink{0000-0002-5936-8921}}
\affiliation{College of Physics and Electronic Information Engineering, Guilin University of Technology, Guilin 541004, China}

\author{Wan-Ting Hou\orcidlink{0009-0004-8194-7446}}
\affiliation{College of Mathematics and Statistics, Liaoning University, Shenyang 110036, China}

\author{Jing-Fei Zhang\orcidlink{0000-0002-3512-2804}}
\affiliation{Liaoning Key Laboratory of Cosmology and Astrophysics, College of Sciences, Northeastern University, Shenyang 110819, China}

\author{Xin Zhang\orcidlink{0000-0002-6029-1933}}\thanks{Corresponding author: zhangxin@neu.edu.cn}
\affiliation{Liaoning Key Laboratory of Cosmology and Astrophysics, College of Sciences, Northeastern University, Shenyang 110819, China}
\affiliation{MOE Key Laboratory of Data Analytics and Optimization for Smart Industry, Northeastern University, Shenyang 110819, China}
\affiliation{National Frontiers Science Center for Industrial Intelligence and Systems Optimization, Northeastern University, Shenyang 110819, China}

\begin{abstract}
Fast radio bursts (FRBs) are commonly classified into repeating and apparently nonrepeating sources, yet whether this distinction reflects intrinsically different physical populations remains uncertain. Using the Second CHIME/FRB Catalog, we apply an unsupervised machine learning framework combining Uniform Manifold Approximation and Projection (UMAP) with density-based clustering to investigate the intrinsic structure of the FRB population in a multi-dimensional parameter space. We find that FRBs are primarily separated into two robust clusters dominated by spectral morphology. One cluster is characterized by narrowband emission and longer durations, while the other exhibits relatively broadband spectra and shorter burst timescales, confirming previous findings based on the first CHIME/FRB catalog. This classification scheme achieves a recall of 0.94 for known repeaters. Within the repeating population, we further identify a stable subclass of atypical repeaters that are broadband, shorter in duration, and more luminous, resembling nonrepeating bursts. Furthermore, broadband nonrepeaters exhibit systematically higher dispersion measures (by approximately 200 $\text{pc cm}^{-3}$) and isotropic luminosities approximately an order of magnitude larger than those of repeating FRBs. Without invoking catastrophic progenitor scenarios, these differences are naturally explained by instrumental sensitivity limits and distance-dependent selection effects. Our results provide new statistical evidence for a physical connection between repeating and nonrepeating FRBs.
\end{abstract}
%\pacs{98.70.Dk, 07.05.Mh, 02.50.Sk}
\maketitle

\section{Introduction}\label{sec:Introduction}

Fast Radio Bursts (FRBs) are millisecond-duration, extremely energetic radio transients. Since their discovery in 2007 \cite{2007Sci...318..777L}, their extreme energetics and cosmological origins have made them one of the most active frontiers in astrophysics, although their physical origins have not yet been fully understood \cite{2013Sci...341...53T,2014ApJ...790..101S,2017ApJ...834L...7T,2017ApJ...834L...8M,2021Natur.598..267L,2022A&ARv..30....2P,2023RvMP...95c5005Z,2024ARNPS..74...89Z}. The burst from the Galactic magnetar SGR 1935+2154 \cite{2020Natur.587...59B,2020Natur.587...54C}, however, has demonstrated that magnetars can account for at least a fraction of the population. Owing to their large dispersion measures (DMs), high brightness, and cosmological distances, FRBs have gradually emerged as powerful probes of the intergalactic medium and the cosmic matter distribution \cite{2020Natur.581..391M,2020ApJ...903...83Z,2023SCPMA..6620412Z,2025SCPMA..6880406Z,2025arXiv250706841Z}. In the era of precision cosmology, addressing key questions such as measuring the Hubble constant and probing the nature of dark energy relies on diverse cosmological observations \cite{2020SciBu..65.1340Z,2024SCPMA..6730411S,2025ApJ...985L..44S,2026SCPMA..6940413S,2026SCPMA..6920401J,2024SCPMA..6720412J,2024SCPMA..6719512L,2025SCPMA..6910413L,2024ApJ...976....1L,2025EPJC...85..608L}. In this context, FRBs have already proven to be a novel tool for independently constraining these fundamental parameters \cite{2023SCPMA..6620412Z,2024ChPhL..41k9801W,2025SCPMA..6880406Z}. Over the past nearly two decades, FRB observations have revealed diverse activity: some sources produce multiple bursts (repeaters), whereas others have so far been detected only once (nonrepeaters). Whether this distinction reflects intrinsically different progenitors or is instead driven by observational incompleteness and selection biases remains one of the most debated questions in current FRB research \cite{2023RvMP...95c5005Z}.

Analyzing the observational characteristics of repeating and nonrepeating FRBs is a key approach to understanding their physical origins. Early studies noted that repeating FRBs typically exhibit more complex temporal structures, broader pulse widths, and narrower spectra \cite{Ziggy2021,2021MNRAS.500.2525K,2022RAA....22l4001Z,2023ApJ...947...83C,2025arXiv250714711Z}, whereas nonrepeating FRBs are characterized by short-duration, broadband emission \cite{2021ApJS..257...59C,Ziggy2021,2021MNRAS.500.3275C,2022Univ....8..355Z}. More recent studies have indicated that repeating and nonrepeating FRBs differ most prominently in their spectral morphology \cite{Ziggy2021,Sun2025}. These trends were subsequently confirmed using the larger CHIME/FRB Catalog 2 sample (Catalog 2) \cite{catalog2}. Compared to apparent nonrepeaters, several active repeating FRBs are associated with compact persistent radio sources \cite{2017Natur.541...58C,niu2022,Moroianu:2025bmg,2025A&A...695L..12B}, and show evidence for long-timescale environmental evolution \cite{2025arXiv250715790W,2026SciBu..71...76N} or pronounced Faraday rotation measure variability \cite{Michilli:2018zec,Hilmarsson:2020npc,2026Sci...391..280L}. These properties are primarily identified in sources under long-term monitoring, and their prevalence among the broader FRB population remains uncertain.

From a statistical perspective, analyses of the burst rates for the two classes show that, after correcting for observational time and sensitivity, their distributions do not exhibit a bimodal structure \cite{2023ApJ...947...83C}. This suggests that repeaters and nonrepeaters may not constitute entirely distinct populations, and that nonrepeating FRBs could potentially repeat on longer timescales. Moreover, correlations between burst rate and pulse width and bandwidth have been identified \cite{Sun2025b}, as required by models that reconcile the apparent morphological dichotomy within a continuous population of repeating FRBs \cite{2023ApJ...947...83C}. In addition, no significant differences have yet been found in the host galaxy properties of the two classes \cite{2022AJ....163...69B,2025ApJ...981...24M,2025ApJ...989L..48C,2026MNRAS.545S2144P}, while analyses of their energy distributions indicate that nonrepeating FRBs may simply represent the rare, high-energy tail of the repeating population \cite{2024NatAs...8..337K,2026MNRAS545f1937O}. Taken together, these findings highlight the complexity of the FRB dichotomy and underscore the need to reassess classification in a higher-dimensional, multi-parameter observational space.

In recent years, machine learning has demonstrated unique advantages in the classification of FRBs. Unsupervised methods can perform dimensionality reduction and clustering on multi-dimensional observational parameters without relying on prior labels, thereby revealing the natural clustering and structural differences between repeating and nonrepeating FRBs in parameter space \cite{2022MNRAS.509.1227C,2023MNRAS.522.4342Y,2023MNRAS.519.1823Z,2024ApJ...977..273G,Sun2025,2025ApJ...982...16Q,Liu2025,2025PASP..137l4102M,2026JHEAp..4900449J}. On the other hand, supervised methods leverage information from known repeating FRBs to establish nonlinear mappings between input features and classes in high-dimensional feature spaces \cite{2023MNRAS.518.1629L,2024MNRAS.533.3283S,2025arXiv250906208K,2025arXiv251102634A,2025arXiv251204204L}, enabling the prediction of potential repeating candidates. Compared to supervised approaches, unsupervised methods are more model-independent and do not require prior training, making them particularly well suited for uncovering statistical correlations among multi-dimensional FRB observables. Ref.~\cite{Sun2025} showed that spectral morphology parameters, particularly the spectral running, serve as key features for distinguishing repeating and nonrepeating FRBs, yielding a clear separation between the two classes. Similar conclusions have been independently confirmed in recent studies \cite{2025arXiv250906208K,2025arXiv251102634A}.

Previous FRB classification studies were largely based on early samples such as Catalog 1 and were therefore limited by small sample sizes, particularly for repeating sources, hindering a comprehensive statistical characterization of their properties. The release of Catalog 2 expands the sample size by nearly an order of magnitude, enabling robust tests of earlier results and systematic analyses of repeating FRBs. Using this expanded dataset, we apply unsupervised machine learning methods with multi-dimensional observational parameters to re-examine FRB classification and explore statistical distinctions and connections between repeating and nonrepeating FRBs, thereby providing new statistical insights into the physical origins of FRBs.

The structure of our paper is as follows: In Sec.~\ref{sec:data}, we describe the data set from Catalog 2 used in this study, along with the implementation of the UMAP algorithm and its parameter settings. In Sec.~\ref{sec:resu}, we present the UMAP clustering results for repeating and nonrepeating FRBs from Catalog 2, along with a detailed comparison of parameter distributions for the reclassified sources and an analysis of the distinctions between typical and atypical repeaters. In Sec.~\ref{Unifying}, we demonstrate that the apparent dichotomy between repeating and nonrepeating FRBs can be naturally explained by instrumental sensitivity limits at large distances and limited exposure times, without the need to invoke distinct physical origins. A brief summary and discussion are provided in Sec.~\ref{sec:concl}. In this work, we adopt the cosmological parameters $H_{0}=67.66$ km s$^{-1}$ Mpc$^{-1}$, $\Omega_{\rm b}=0.04897$, $\Omega_{\rm m}=0.3111$, and $\Omega_{\Lambda}=0.6874$ from the {\tt Planck 2018} results \cite{Planck:2018vyg}.

\section{Data and algorithm parameters} \label{sec:data}

\subsection{Data sample selection} \label{subsec:datasel}
This study employs the latest and currently largest FRB dataset released by the CHIME/FRB project, the second CHIME/FRB Catalog \cite{catalog2}. All events are detected by the same instrument operating in the 400--800 MHz band under uniform observational conditions, yielding a highly homogeneous sample of FRB events. Such a consistent and statistically rich dataset provides an unprecedented foundation for population studies. Catalog 2 covers 4539 FRBs detected between 2018 July 25 and 2023 September 15 \cite{catalog2}. Many of these FRBs exhibit complex temporal and spectral structures and can be decomposed into multiple sub-bursts, which appear as distinct peaks in the dynamic spectrum \cite{2024MNRAS.52710425S,2024MNRAS.529L.152B}. Decomposing such multi-component bursts yields a total of 5045 sub-bursts. Since these sub-components often differ in spectral properties, we treat each sub-burst as an independent burst in our analysis. We exclude 518 FRBs that lack flux measurements. Notably, this criterion inherently filters out all events flagged as far sidelobe detections. Our final dataset includes 4527 FRBs in total, consisting of 3373 nonrepeaters and 1154 repeaters originating from 83 distinct repeating sources.

\subsection{FRB feature parameters} \label{ssec:data desc}

In our previous studies \cite{Sun2025,Sun2025b}, we provided a detailed description of the characteristic parameters that reflect the intrinsic emission properties and spectral characteristics of FRBs, which are capable of distinguishing between repeating and nonrepeating FRBs. Our goal was to identify the most critical and minimal set of parameters needed to differentiate repeating FRBs from nonrepeating ones. To achieve this, we selected a small number of directly observed parameters and applied machine learning classification techniques for the analysis. The chosen parameters include flux density ($S_{\nu}$), fluence ($F_{\nu}$), sub-burst pulse width ($\Delta t_\mathrm{WS}$), spectral index ($\gamma$), spectral running ($r$), the lowest frequency ($\nu_\mathrm{Low}$), the highest frequency ($\nu_\mathrm{High}$), and peak frequency ($\nu_\mathrm{Peak}$). These parameters have been shown to be effective for distinguishing repeaters from nonrepeaters when used with machine learning techniques. 

The spectral morphology of each burst is characterized using the empirical spectral model defined in Catalog 2 \cite{catalog2}. In this model, the spectral index $\gamma$ and the spectral running parameter, denoted here as $r$ (referred to as $\beta$ in Catalog 2), describe the burst spectrum through:
\begin{equation}
S(\nu) = A \left( \frac{\nu}{\nu_0} \right)^{\gamma + r \ln \left( \frac{\nu}{\nu_0} \right)},
\label{equ:gammar}
\end{equation}
where $A$ is the amplitude, and $\nu_0 = 600\,{\rm MHz}$ is the reference frequency used in Catalog 2 fits. While $\gamma$ describes the instantaneous spectral slope at $\nu_0$, the running parameter $r$ captures the spectral curvature. A key update in Catalog 2 is that $r$ is allowed to take large negative values, enabling this single functional form to model both broadband spectra ($r \approx 0$) and narrowband Gaussian-like structures ($r \ll 0$).

This study considers eight parameters from Catalog 2: $S_{\nu}$, $F_{\nu}$, $\Delta t_\mathrm{WS}$, $\gamma$, $r$, $\nu_\mathrm{Low}$, $\nu_\mathrm{High}$, and $\nu_\mathrm{Peak}$. Detailed definitions of these parameters are provided in Ref.~\cite{catalog2}. Prior to dimensionality reduction, all parameters except $\gamma$ and $r$ were log-transformed; subsequently, all eight features were Z-score standardized.

\subsection{UMAP algorithm} \label{ssec:algor}
To investigate potential structures and similarity relationships among FRBs in the multi-dimensional parameter space, we employ the Uniform Manifold Approximation and Projection (UMAP) algorithm \cite{McInnes:2018dzu,becht2019dimensionality}. As a nonlinear dimensionality-reduction technique grounded in Riemannian geometry and algebraic topology, UMAP assumes that the data can be approximated as lying on a locally connected manifold. The algorithm constructs a high-dimensional fuzzy topological representation and optimizes the low-dimensional embedding by minimizing the cross-entropy between the high-dimensional and low-dimensional topological graphs. A distinct advantage of UMAP over similar methods, such as t-SNE \cite{JMLR:v9:vandermaaten08a}, is its higher computational efficiency on large datasets and its ability to preserve the global structure of the data more effectively while simultaneously maintaining local neighborhood relationships. This makes it highly suitable for revealing meaningful substructures or clustering tendencies within complex astrophysical datasets.

The topology of the resulting embedding is primarily governed by two key hyperparameters: n\_neighbors and min\_dist. The n\_neighbors parameter determines the size of the local neighborhood used to approximate the manifold structure and serves a role analogous to perplexity in t-SNE, balancing the emphasis between fine local details and the broader global structure. The min\_dist parameter controls the minimum separation distance between points in the low-dimensional space. Lower values allow points to clump tightly together, potentially revealing denser substructures, whereas higher values lead to a more dispersed distribution, which aids in visualizing the continuous topological structure. Proper tuning of these parameters is crucial for generating a stable and physically interpretable embedding. Throughout this analysis, we adopt n\_neighbors = 15 and min\_dist = 0.1 for UMAP dimensionality reduction. We find that this commonly used setting provides a good balance between preserving local fine structure and maintaining global topology. The robustness of the embedding against variations in n\_neighbors is quantitatively assessed in Sec.~\ref{ssec:stability}. All other hyperparameters are kept at their default values as specified in the umap-learn library. Further details on the UMAP implementation and parameter tuning can be found in the official documentation\footnote{\url{https://umap-learn.readthedocs.io/}}.

\section{Results and discussion} \label{sec:resu}

\subsection{UMAP results} \label{subsec:class}

\begin{figure*}[ht!]
\centering
\includegraphics[angle=0,scale=0.4]{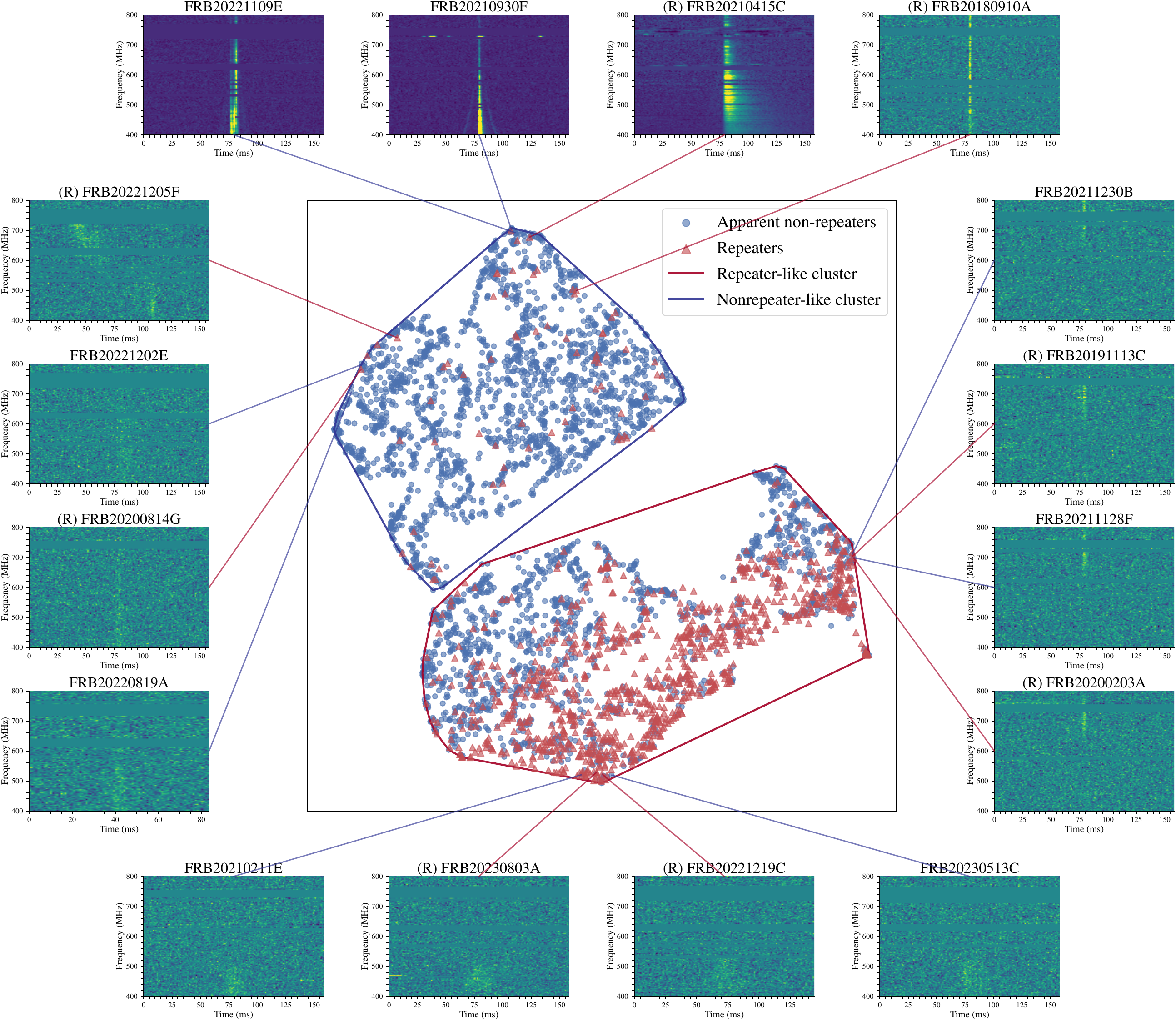}
\caption{Two-dimensional UMAP projection of FRBs in the Catalog 2 sample, with clusters identified by HDBSCAN. Apparent nonrepeaters and repeaters are shown as blue circles and red triangles, respectively. Blue and red contours delineate the HDBSCAN-identified nonrepeater-like and repeater-like clusters. The surrounding panels display dynamic spectra of representative FRBs from each cluster, with red and blue lines pointing to repeaters and apparent nonrepeaters, respectively.
\label{fig:figure1}}
\end{figure*}

To visualize the distribution of 4527 FRBs in the multidimensional parameter space, we apply UMAP to the sample, and the result is shown in Figure~\ref{fig:figure1}. In this embedding space, red triangles represent confirmed repeaters, while blue circles denote FRBs that have not yet been observed to repeat. Consistent with the foundational patterns identified in Refs.~\cite{Sun2025,Sun2025b}, the enlarged sample still exhibits an approximately bimodal structure. As illustrated, most confirmed repeaters are densely concentrated on the lower-right side. In contrast, the distribution of apparent nonrepeaters is more complex: while a significant portion forms a distinct, isolated cluster in the upper-left region, others are widely distributed and show substantial spatial overlap with the primary repeaters in the lower-right. It is worth noting that, in the upper-left region of the embedding space, a subset of repeaters now appears within areas dominated by nonrepeaters, which constitutes a notable change relative to earlier results derived from the smaller Catalog 1 sample \cite{Sun2025}.

It should be noted that the UMAP axes themselves do not carry direct physical meaning, and distances in the embedded space merely reflect relative similarities in the multidimensional parameter space. Nevertheless, the projection reveals a clear global trend of separation between repeaters and nonrepeaters, indicating that the two populations remain systematically distinct in their parameter properties even in the enlarged sample. These distributional characteristics motivate the use of objective clustering methods to further dissect the underlying populations and facilitate a more physically meaningful interpretation of the observed patterns.

To this end, we apply the Hierarchical Density-Based Spatial Clustering of Applications with Noise (HDBSCAN) algorithm \cite{McInnes2017} to the UMAP embedded space. This method does not require a predefined number of clusters, is capable of identifying groups of arbitrary shape and size in complex distributions, and can exclude noise points. As shown in Figure~\ref{fig:figure1}, HDBSCAN partitions the sample into two major clusters, delineated by red and blue contours. The lower cluster contains the vast majority of confirmed repeaters, which we designate as the repeater-like cluster, while the upper cluster is dominated by apparent nonrepeaters, referred to as the nonrepeater-like cluster. Notably, a clear overlap is present between the two clusters: the repeater-like cluster includes many FRBs that have not yet been observed to repeat, while the nonrepeater-like cluster contains a small number of repeating bursts.

To aid the interpretation of the HDBSCAN clustering results and the UMAP embedding, we randomly select representative FRBs from each category and display their dynamic spectra (waterfall plots). These visual examples allow a direct comparison of their spectral-temporal features and overall morphology, providing complementary insight into the statistical classification results.

From the comparison of the waterfall plots of repeaters and nonrepeaters shown in the lower panels of Figure~\ref{fig:figure1}, it is evident that the apparent nonrepeaters residing within the repeater-like cluster exhibit strong morphological similarities to confirmed repeaters. In particular, they share a dominance of narrowband, relatively faint emission features with long pulse widths. Such similarities suggest that these apparent nonrepeaters may in fact be latent repeaters, whose subsequent bursts have not yet been detected. However, as noted in Ref.~\cite{catalog2}, identifying genuine candidates from this population in the future will require improved localization to rigorously exclude instrumental artifacts (e.g., grating-lobe detections) that can mimic repeater-like features.

For repeaters located in the nonrepeater-like cluster, inspection of the waterfall plots in the upper panels of Figure~\ref{fig:figure1} shows that these bursts are predominantly broadband, short in pulse width, and comparatively bright, closely resembling the spectral morphology typically observed in apparent nonrepeating bursts. The presence of confirmed repeaters in the nonrepeater-like cluster demonstrates that repeating sources can occasionally display observational properties commonly attributed to nonrepeaters. Such an overlap in the high-dimensional feature space indicates that the distinction between repeaters and nonrepeaters is not strictly dichotomous. Instead, it supports the interpretation that a substantial fraction--if not all--FRBs may arise from a common physical origin or represent different manifestations along an activity sequence. Hereafter, we refer to repeating FRBs that fall within the nonrepeater-like cluster as ``atypical repeaters'', while those located in the repeater-like cluster are referred to as ``typical repeaters''.

\begin{figure}[ht!]
\centering
\includegraphics[angle=0,scale=0.5]{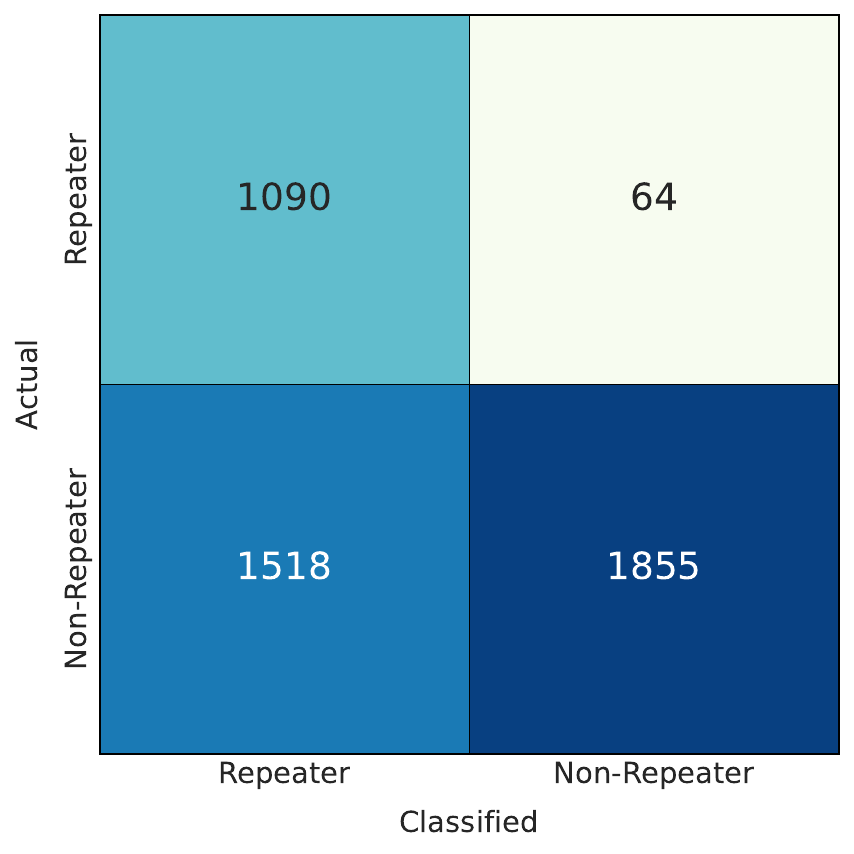}
\caption{Confusion matrix of the clustering classification.
\label{fig:figure2}}
\end{figure}

Based on the UMAP embedding and HDBSCAN clustering results, we quantitatively assess the performance of the unsupervised framework in recovering known repeating FRBs. Figure~\ref{fig:figure2} presents the classification confusion matrix. Confirmed repeaters assigned to the repeater-like cluster are counted as correctly classified. For the full sample of 4527 FRBs, 1090 repeaters are correctly identified by unsupervised framework. The recall is defined as:
\begin{equation}
{\rm Recall} = \frac{T_{\rm P}}{T_{\rm P} + F_{\rm N}},
\label{equ:recall}
\end{equation}
where $T_{\rm P}$ (true positives) denotes the number of repeaters correctly classified into the repeater-like cluster, and $F_{\rm N}$ (false negatives) represents repeaters that are incorrectly assigned to the nonrepeater-like cluster. Under this definition, we obtain a recall of 0.94, indicating that the majority of known repeating FRBs are grouped into the repeater-like cluster.

This high recall demonstrates that the model remains robust in capturing the fundamental high-dimensional characteristics of repeaters, even with a significantly enlarged sample. Furthermore, the confusion matrix shows that 1518 apparent nonrepeaters are assigned to the repeater-like cluster. However, this does not indicate a failure of the classification. Instead, as illustrated by their waterfall plots, these sources exhibit morphological properties--such as narrowband emission and long pulse widths--that are indistinguishable from those of known repeaters. This aligns with the hypothesis that these events likely represent a population of latent repeaters \cite{Sun2025,2024MNRAS.52711158Y,2025arXiv250906208K}, while cautioning that instrumental effects should be carefully considered. In our earlier analysis of Catalog 1 using a similar unsupervised framework, we identified a group of repeater candidates \cite{Sun2025}, two of which---FRB 20190208C and FRB 20190329A---have since been confirmed as repeating sources in Catalog 2. Consequently, the high recall and the structured distribution observed in the UMAP projection confirm that our framework captures physically meaningful properties of FRB populations, providing a basis for identifying repeater candidates in large-scale surveys.

It is worth noting that the aforementioned recall of 0.94 is derived from a burst-level analysis. As discussed in Section~\ref{ssec:distribution}, since bursts from repeating sources tend to exhibit narrowband characteristics, this statistical correlation may inflate the evaluated recall. To analyze this potential statistical difference and verify the robustness of our clustering, we conduct a source-level analysis, representing each of the 83 distinct repeating sources solely by the first sub-burst of its earliest observed event. Interestingly, when evaluating these alongside 3190 nonrepeating sources, the unsupervised model's recall for repeaters decreases to 0.72 (details in Appendix A). Furthermore, the decrease in recall clearly highlights that individual bursts from known repeaters exhibit substantial morphological diversity. To comprehensively characterize the FRB parameter space and preserve all available burst information, subsequent analyses in this paper will proceed utilizing the burst-level framework.

\subsection{Atypical repeaters and the blurred boundary of FRB populations} \label{subsec:atypical}

Interestingly, as shown in Figure~\ref{fig:figure1}, 64 confirmed repeaters are assigned to the nonrepeater-like cluster. In our initial analysis based on Catalog 1, only a single repeating FRB was assigned to the nonrepeater-like cluster, which was insufficient to draw firm conclusions \cite{Sun2025}. With the subsequent increase in the number of known repeaters, our follow-up study incorporating both Catalog 1 and the CHIME/FRB Catalog 2023 sample revealed that a small but non-negligible fraction of repeaters consistently occupies the nonrepeater-like cluster \cite{Sun2025b}. In that analysis, atypical repeating FRBs accounted for 4.9\% of the nonrepeater-like cluster and approximately 12\% of the repeating FRB population, suggesting the possible emergence of a distinct subclass.

With the significantly expanded and more statistically robust Catalog 2 sample, we find that the absolute number of repeaters assigned to the nonrepeater-like cluster continues to increase. However, due to the substantial growth of the overall sample, their fractional contribution to the nonrepeater-like cluster decreases to 3.3\%, while their fraction among all repeating FRBs is measured to be 6\%. Although these fractions are lower than those reported in Ref.~\cite{Sun2025b}, this difference is likely attributable to the limited statistical sample of repeating FRBs in the earlier analysis. Regardless, within the significantly enlarged Catalog 2 sample, the persistent occupancy of a subset of atypical repeaters in the nonrepeater-like cluster remains a robust and statistically significant finding. This sustained segregation confirms the existence of a stable and distinct subclass of repeaters whose observed characteristics are indistinguishable from typical nonrepeating FRBs.

\begin{figure}[ht!]
\centering
\includegraphics[angle=0,scale=0.42]{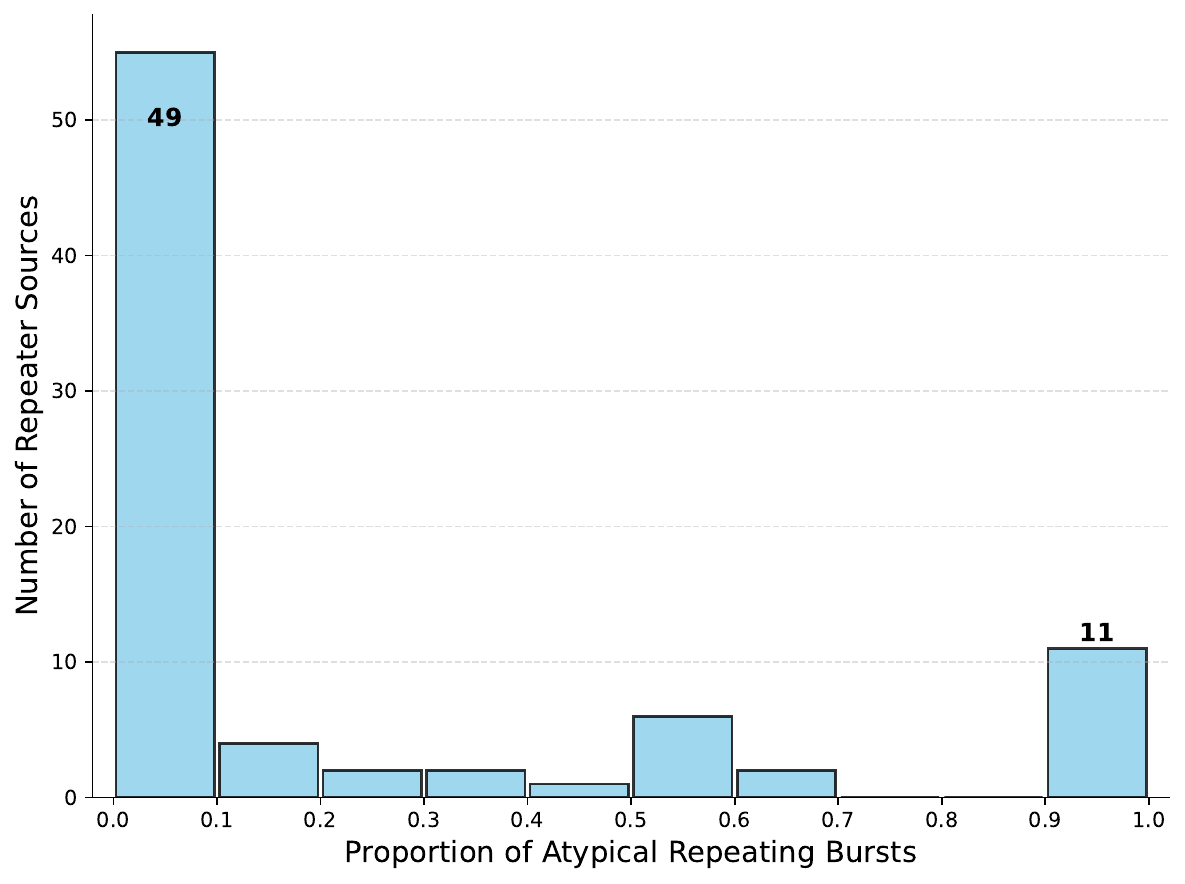}
\caption{Distribution of the proportion of atypical repeating bursts within the repeater sources.
\label{fig:figure3}}
\end{figure}

To further quantify the prevalence of atypical repeaters within repeating FRB sources, we examine the fraction of atypical bursts produced by each individual repeating source. As shown in Figure~\ref{fig:figure3}, among the 83 known repeating FRB sources, 49 exhibit exclusively typical repeating bursts. Of the remaining sources, 23 display mixed behavior with both typical and atypical bursts, while 11 exhibit exclusively atypical repeating bursts (atypical fraction = 1.0). Although these 11 sources generally have low burst counts (2-3 detected bursts), which may reflect statistical fluctuations, their presence nevertheless indicates a subset of repeaters whose observed properties are indistinguishable from apparent one-off events.

The detection of atypical broadband bursts within the repeating FRB population has significantly blurred the phenomenological boundary between repeating and apparently nonrepeating FRBs. While repeaters are often characterized by narrow spectral envelopes, the presence of broadband events indicates that their spectral properties span a wider range than previously assumed. The narrow spectra frequently observed in repeaters therefore likely arise from an intrinsic emission mechanism \cite{2023ApJ...956...67Y,2024ApJ...974..160K,2024A&A...685A..87W}, rather than propagation effects (e.g., interstellar or intergalactic plasma lensing and absorption). Nevertheless, it is worth noting that localized propagation effects within the immediate source environment, such as a dense magnetar wind nebula, cannot be fully excluded and may still induce spectral modulation. In addition, polarization measurements from repeating FRBs and the only known FRB-associated magnetar suggest that their progenitors likely possess dynamically evolving magnetospheres \cite{2023SciA....9F6198Z,2025ApJ...988..175L}. In such a dynamic environment, bursts may originate from different emission heights along the magnetic field lines, where the magnetic geometry varies with altitude. This naturally leads to burst-to-burst diversity in observed properties, including spectral bandwidth, central frequency, and polarization behavior. Such a physical picture is consistent with the empirical observation that repeaters predominantly exhibit narrowband spectra but occasionally display a broadband, nonrepeater-like morphology, and it is this observed morphological overlap that implies a tighter physical link between these two observed classes than previously thought. The coexistence of narrowband and broadband emission within the same magnetar \cite{2026arXiv260321450L} further supports this picture. These findings suggest that traditional classifications based on single-burst morphology may be tracing different emission conditions rather than fundamentally distinct progenitors.

\subsection{Dimensionality reduction stability} \label{ssec:stability}

The UMAP embedding is primarily influenced by the number of nearest neighbors (n\_neighbors). To ensure that the observed bimodality in the FRB population is physically meaningful rather than an artifact of specific parameter tuning, we perform a stability test by varying n\_neighbors from 5 to 25.

\begin{figure*}[htbp]
\centering
\includegraphics[width=1.0\textwidth]{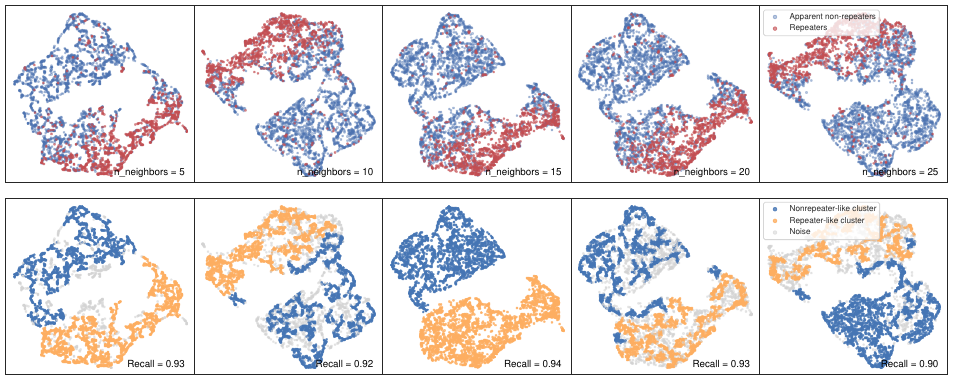}
\caption{UMAP embedding stability (upper panel) and HDBSCAN clustering results (lower panel) for different n\_neighbors values.}
\label{fig:figure4}
\end{figure*}

As illustrated in the upper panel of Figure~\ref{fig:figure4}, despite variations in local density, the global separation between the repeating (red) and nonrepeating (blue) populations remains conspicuous and topologically consistent across all tested values. To quantify the performance of the unsupervised results, we apply HDBSCAN clustering to each projection and assess the clustering outcome using the recall metric defined in Equation~\ref{equ:recall}. The lower panel of Figure~\ref{fig:figure4} demonstrates that the clustering results are highly robust: the repeater-like cluster successfully recover the majority of confirmed repeaters. The recall scores consistently exceed 0.90 throughout the hyperparameter range, confirming that the classification of FRBs based on this embedding is statistically stable and insensitive to minor hyperparameter variations.

Although the classification performance remains robust across the tested n\_neighbors range, we observe variations in the noise fraction and boundary sharpness. Consequently, n\_neighbors = 15 is adopted as the reference parameter for the final analysis, as it effectively preserves both local and global data topology and yields the maximum recall of 0.94, thereby maximizing the recovery of repeating FRBs. In the HDBSCAN clustering process, we systematically evaluate parameter combinations of min\_cluster\_size (50--500) and min\_samples (2--300) using the Density-Based Clustering Validation (DBCV) score \cite{Moulavi2014}. The average DBCV score exceeds 0.44 across most configurations, confirming robust topological separation and predominantly yielding a stable two-cluster structure. While the algorithm can be tuned to identify more sub-clusters, such finer subdivision yields no actual improvement in repeater recall. We ultimately adopt min\_cluster\_size = 100 and min\_samples = 5. This configuration achieves a high DBCV score of 0.53 with zero noise points, ensuring a physically interpretable partition of the FRB sample.

\subsection{Feature importance} \label{ssec:feature}

To further elucidate the discriminative capability of the machine learning model in separating repeating and nonrepeating FRBs, we perform a systematic analysis of input feature importance using the SHAP (SHapley Additive exPlanations; \citealt{Lundberg2017}) method. Based on the Shapley value concept from game theory, SHAP quantifies the marginal contribution of each feature to the model's output, thereby revealing its significance at both the global and individual levels. Compared with traditional feature importance metrics, SHAP yields more stable and interpretable results, specifically by differentiating how various feature values influence the prediction outcomes.

\begin{figure}[ht!]
\centering
\includegraphics[angle=0,scale=0.58]{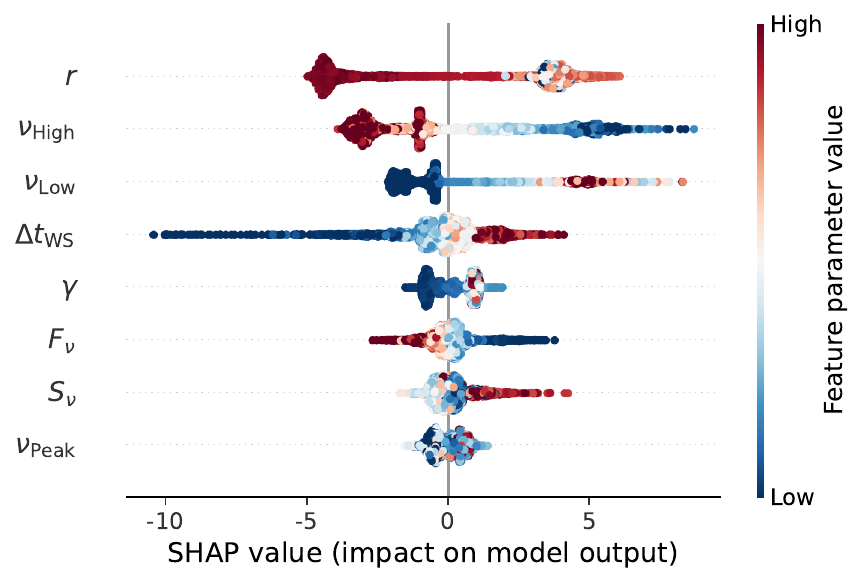}
\caption{Feature importance and impact distribution derived from SHAP analysis. Features are ordered vertically by decreasing global importance. Colors encode the feature values (red to blue), while the horizontal axis indicates the direction and magnitude of each feature’s contribution to the model output.
\label{fig:figure5}}
\end{figure}

In this study, we implement the SHAP framework in conjunction with a supervised XGBoost classifier \cite{chen2016xgboost} and calculate SHAP values for all input parameters, summarizing their distributions, as shown in Figure~\ref{fig:figure5}. In this plot, features are arranged along the vertical axis in descending order of overall importance. The horizontal axis represents the magnitude and sign of SHAP values, indicating both the direction and strength of each feature's contribution to the classification. The color scale, ranging from blue to red, corresponds to increasing feature values. During model training, the repeater class is treated as the positive class. Consequently, positive SHAP values indicate that a given feature value contributes positively toward classifying an FRB as a repeater, whereas negative values favor classification as a nonrepeater.

The SHAP analysis in Figure~\ref{fig:figure5} elucidates the underlying decision-making process, confirming that the parameter $r$ remains the most influential feature in the enlarged Catalog 2 sample, consistent with our prior findings \cite{Sun2025,Sun2025b}. More negative values of $r$ tend to push the model toward classifying FRBs as repeaters, typically corresponding to steep, narrowband bursts, whereas smaller absolute values of $r$ are more often associated with nonrepeaters, reflecting flatter, broadband emission. Representative waterfall plots in Figure~\ref{fig:figure1} visually validate this pattern.

Importantly, the consistently dominant influence of $r$ across different datasets demonstrates the internal robustness and stability of our classification framework \cite{Sun2025,Sun2025b}. This consistency ensures that variations in the inferred fraction of atypical repeaters across different sample selections are directly comparable, providing confidence in the analysis and interpretation of their relative occurrence rates described above.

Beyond spectral morphology parameters, frequency and temporal parameters also play critical but secondary roles. SHAP value distributions indicate that spectral bandwidth contributes to classification, with lower $\nu_\mathrm{High}$ and higher $\nu_\mathrm{Low}$ (narrower bandwidth emission) being more strongly associated with repeaters. The temporal width $\Delta t_{\mathrm{WS}}$ has a distinct bifurcated effect: larger burst widths favor classification as repeaters, whereas narrow pulses favor the nonrepeater-like cluster. While the spectral index $\gamma$ remains a theoretically important parameter, its impact in the current model appears partially overshadowed by the dominant diagnostic power of $r$. In contrast, flux-related parameters ($F_\nu$ and $S_\nu$) and the peak frequency $\nu_\mathrm{Peak}$ are less important overall, though they still provide supplementary information for classifying a subset of events. Notably, nonrepeaters are generally associated with higher $F_\nu$ values.

\subsection{Parameter distributions} \label{ssec:distribution}

In Sec.~\ref{subsec:class}, the Catalog 2 sample is separated into two primary groups using an unsupervised framework that combines UMAP dimensionality reduction with the HDBSCAN clustering algorithm. Feature-importance analyses and comparisons of representative burst dynamic spectra suggest that this separation reflects the presence of two distinct spectral-morphological regimes within the FRB population. However, it remains unclear whether these observational regimes reflect fundamentally different progenitor classes, or are more plausibly attributable to distinct radiative emission mechanisms. In this section, we perform a detailed statistical analysis of the parameter distributions to quantify and compare the observational differences between FRBs belonging to these two groups.

To investigate the intrinsic properties of FRBs, we further examine the distributions of their isotropic energy ($E_{\rm iso}$), isotropic peak luminosity ($L_{\rm iso}$), and brightness temperature ($T_{\rm B}$). Because the majority of FRBs remain without host identifications and thus lack spectroscopic redshift measurements, we estimate the redshift $z$ from DMs using the Catalog 2 as the reference. The total observed DM can be decomposed into several physically distinct components \cite{2023MNRAS.519.1823Z}:
\begin{equation}
{\rm DM} = {\rm DM_{MW}} + {\rm DM_{Halo}} + {\rm DM_{IGM}} + \frac{{\rm DM_{Host}}}{1+z},
\end{equation}
where $\rm DM_{MW}$, $\rm DM_{Halo}$, $\rm DM_{IGM}$, and $\rm DM_{Host}$ represent the respective contributions from the Milky Way interstellar medium, the Milky Way halo, the intergalactic medium, and the host galaxy of the FRB. We compute $\rm DM_{MW}$ using the NE2001 model \cite{2002astro.ph..7156C}. Consequently, the extragalactic component is given by
\begin{equation}
{\rm DM_E} = {\rm DM} - {\rm DM_{MW}} - {\rm DM_{Halo}} = {\rm DM_{IGM}} + \frac{{\rm DM_{Host}}}{1+z}.
\end{equation}

Following commonly adopted assumptions in the literature, we take ${\rm DM_{Halo}} = 30\ \mathrm{pc\ cm^{-3}}$ \cite{2015MNRAS.451.4277D, 2021MNRAS.501.5319A, 2023MNRAS.519.1823Z}, and adopt ${\rm DM_{Host}} = 130\ \mathrm{pc\ cm^{-3}}$ based on recent estimates from IllustrisTNG simulations \cite{2024arXiv240916952C}. 

In the standard $\Lambda$CDM cosmology, 
the mean IGM contribution can be written as \cite{2014ApJ...783L..35D,2014ApJ...788..189G,2020Natur.581..391M}:
\begin{equation}
\label{eq:dmigm}
\langle{\rm DM_{IGM}}(z)\rangle =
\frac{3 c H_0 \Omega_{\rm b}}{8\pi G m_{\rm p}}
\int_{0}^{z} 
\frac{f_{\rm d}(z)\, f_{\rm e}(z)\, (1+z)}
{\sqrt{\Omega_{\rm m}(1+z)^{3}+\Omega_{\Lambda}}}\, dz ,
\end{equation}
where $m_{\rm p}$ is the proton mass. The functions $f_{\rm d}(z)$ and $f_{\rm e}(z)$ describe the baryon mass fraction residing in the IGM and the ionized electron fraction per baryon, respectively. In this work, we adopt the recent constraints from Ref.~\cite{2024arXiv240916952C}: $f_{\rm d} = 0.93$, $f_{\rm e} = 0.875$, and $\Omega_{\rm b} h_{70} = 0.049$. Cosmological parameters follow the {\tt Planck 2018} results, as summarized in the introduction. For sources with small $\rm DM_{E}$, the host-galaxy contribution can exceed that of the IGM, introducing large uncertainties in the inferred redshift. To reduce this bias, our redshift estimation is restricted to FRBs with $\rm DM_{E} > 160\ pc\ cm^{-3}$.

With the estimated redshifts, we proceed to compute $E_{\rm{iso}}$, $L_{\rm{iso}}$, and $T_{\rm{B}}$ for each FRB. Following Ref.~\cite{2018ApJ...867L..21Z}, $E_{\rm{iso}}$ can be written as:
\begin{equation}
\begin{aligned}
E_{\rm{iso}} & \simeq \frac{4 \pi D_{\mathrm{L}}^{2}}{(1+z)} F_{\mathrm{obs}} \nu \\
&=\left(10^{39} \mathrm{erg}\right) \frac{4 \pi}{(1+z)}\left(\frac{D_{\mathrm{L}}}{10^{28} \mathrm{~cm}}\right)^{2} \frac{F_{\mathrm{obs}}}{\mathrm{Jy} \cdot \mathrm{ms}} \frac{\nu}{\mathrm{GHz}},
\end{aligned}
\end{equation}
and $L_{\rm{iso}}$ is given by
\begin{equation}
\begin{aligned}
L_{\rm{iso}} & \simeq 4 \pi D_{\mathrm{L}}^{2} S_{\nu, \mathrm{p}} \nu\\
&=\left(10^{42} \operatorname{erg~s}^{-1}\right) 4 \pi\left(\frac{D_{\mathrm{L}}}{10^{28} \mathrm{~cm}}\right)^{2} \frac{S_{\nu, \mathrm{p}}}{\mathrm{Jy}} \frac{\nu}{\mathrm{GHz}},
\end{aligned}
\end{equation}
where $F_{\rm obs}$ denotes the measured fluence (expressed in Jy$\cdot$ms, equivalent to erg~cm$^{-2}$~Hz$^{-1}$), $S_{\nu,\mathrm{p}}$ is the peak flux (in units of erg~s$^{-1}$~cm$^{-2}$~Hz$^{-1}$ or Jy), and $D_{\rm L}$ is the luminosity distance.

The brightness temperature $T_{\rm{B}}$ is calculated using the expression adopted in Refs.~\cite{2019A&ARv..27....4P, 2022A&A...657L...7X}:
\begin{equation}
\begin{aligned}
T_{\rm B}
& \simeq \frac{S_{\nu, \mathrm{p}} D_{\mathrm{L}}^{2}}
{2 \pi k_{\mathrm{B}}(\nu \Delta t)^{2}(1+z)} \\
& = 1.1 \times 10^{35}\,\mathrm{K}
\left(\frac{S_{\nu, \mathrm{p}}}{\mathrm{Jy}}\right)
\left(\frac{D_{\mathrm{L}}}{\mathrm{Gpc}}\right)^{2}
\left(\frac{\nu}{\mathrm{GHz}}\right)^{-2}
\left(\frac{\Delta t}{\mathrm{ms}}\right)^{-2} \\
& \quad \times \frac{1}{1+z}.
\end{aligned}
\end{equation}
where $k_{\mathrm{B}}$ is the Boltzmann constant and $\Delta t$ is the burst duration. In this work, we follow the convention adopted in the Catalog 2 and evaluate $E_{\rm iso}$, $L_{\rm iso}$, and $T_{\rm B}$ using a uniform observational bandwidth of $\nu = 400$ MHz corresponding to the CHIME observing band, rather than the central frequency adopted in Ref.~\cite{2018ApJ...867L..21Z}.

In addition to the energy quantities, the parameter distribution analysis presented below focuses on a subset of parameters---specifically $S_{\nu}$, $F_{\nu}$, $\Delta t_\mathrm{WS}$, $\gamma$, and $r$---which are most directly related to burst temporal properties and spectral morphology. We further include the frequency bandwidth $\Delta\nu$, together with three additional parameters defined in Ref.~\cite{catalog2}: the scattering time $\Delta t_\mathrm{ST}$, the dispersion measure DM, and the extragalactic dispersion measure eDM, in order to provide a more comprehensive comparison of parameter distributions among different FRB populations.

\subsubsection{Parameter comparison of repeater-like and nonrepeater-like clusters} \label{sssec:distribution_Nonrepeater}

\begin{figure*}[t]
\centering
\includegraphics[width=1.0\textwidth]{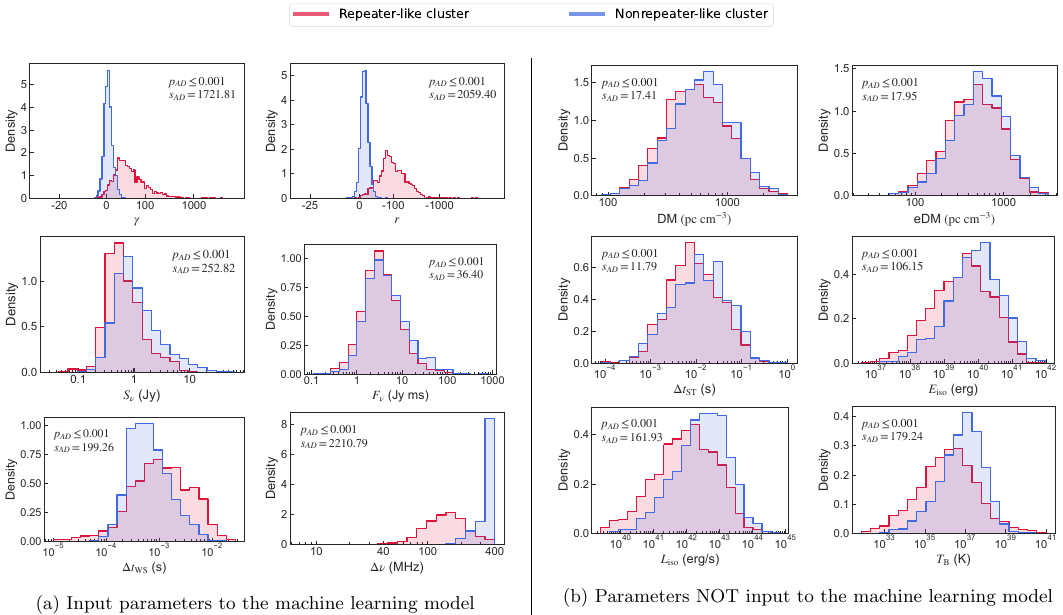}
\vspace{-1ex}
\caption{Parameter distribution comparison between the repeater-like and nonrepeater-like clusters. Panel (a) displays the physical parameters utilized as inputs for the UMAP and HDBSCAN classification. Note that although $\Delta \nu$ was not directly input to the model, it is presented here because it is derived from the actual input parameters $\nu_{\rm Low}$ and $\nu_{\rm High}$ ($\Delta \nu = \nu_{\rm High} - \nu_{\rm Low}$). Panel (b) presents the distributions of the parameters not input to the machine learning model. The $p_{\rm AD}$ and $s_{\rm AD}$ values in each panel denote the $p$-value and test statistic of the AD test, respectively, quantifying the statistical significance of the distributional differences between the two clusters. Note that $p$-values are limited at a minimum of 0.001.}
\vspace{-2ex}
\label{fig:figure6}
\end{figure*}

To comprehensively characterize the parameter differences within the reclassified Catalog 2 sample, we temporarily set aside the repeater/nonrepeater labels and instead perform a systematic comparison of the parameter distributions of FRBs belonging to the repeater-like and nonrepeater-like clusters. We adopt the Anderson-Darling (AD) test \cite{anderson1954test} to quantitatively assess the statistical divergence between the two populations. The AD test evaluates the null hypothesis that the two samples are drawn from the same parent distribution. A $p$-value smaller than 0.05 leads to rejection of the null hypothesis, indicating a statistically significant difference between the repeater-like and nonrepeater-like clusters. Figure~\ref{fig:figure6} presents the resulting distributions together with the corresponding AD test statistics. In particular, $p_{\mathrm{AD}} \leq 0.001$ indicates significant differences across the examined parameters, while the normalized test statistic $s_{\mathrm{AD}}$ provides a direct measure of the strength of the distributional separation.

As shown in Figure~\ref{fig:figure6}, a global statistical comparison reveals that the repeater-like cluster (red) and the nonrepeater-like cluster (blue) differ significantly across all examined physical parameters ($p_{\mathrm{AD}} \leq 0.001$). However, an important caveat should be considered regarding the interpretation of these $p$-values. In large-sample analyses, the AD test becomes highly sensitive, amplifying even minor deviations from the null hypothesis. Therefore, although these $p$-values indicate that the distributions are not strictly identical, the physical significance of the difference is better measured by the magnitude of $s_{\mathrm{AD}}$. The most pronounced separation arises from the spectral morphology parameters $r$ ($s_{\mathrm{AD}} = 2059.40$) and $\gamma$ ($s_{\mathrm{AD}} = 1721.81$), whose exceptionally large normalized AD statistics indicate that spectral morphology plays a dominant role in driving the segregation observed in the high-dimensional feature space.

A detailed statistical summary of these parameters is presented in Table~\ref{tab:tab1}. Given that the distributions of many physical parameters deviate from a Gaussian form, we adopt median values, with uncertainties estimated via bootstrap resampling, as a robust statistical measure to characterize and compare the two clusters.

\begin{table*}[ht!]
\footnotesize %\scriptsize
\centering
\caption{Comparison of median parameters across different FRB classification groups. Median values and their associated uncertainties are calculated via bootstrap resampling. ``Narrowband Nonrep.'' and ``Broadband Nonrep.'' refer to nonrepeaters identified within the repeater-like and nonrepeater-like clusters, respectively.}
\label{tab:tab1}
\begin{tabular*}{\textwidth}{@{\extracolsep{\fill}} l cc cc ccc @{}}
\hline \hline
Feature & \multicolumn{2}{c}{Cluster-based} & \multicolumn{2}{c}{Repeater Subtypes} & \multicolumn{3}{c}{Primary Contrast} \\
\cline{2-3} \cline{4-5} \cline{6-8}
& Rep.-like & Nonrep.-like & Typical Rep. & Atypical Rep. & Narrowband Nonrep. & Broadband Nonrep. & All Rep. \\
\hline
$\gamma$                               & $33.5 \pm 1.0$    & $1.98 \pm 0.14$  & $54.1 \pm 1.5$    & $3.9 \pm 1.1$    & $22.3 \pm 0.7$ & $1.92 \pm 0.14$  & $50.6 \pm 1.4$    \\
$r$                                    & $-88.6 \pm 1.6$   & $-6.66 \pm 0.19$ & $-140.1 \pm 3.7$  & $-9.6 \pm 1.2$   & $-69.5 \pm 1.1$ & $-6.56 \pm 0.18$ & $-133.0 \pm 3.6$  \\
$S_{\nu}$ (Jy)                         & $0.590 \pm 0.010$ & $0.926 \pm 0.019$ & $0.532 \pm 0.013$ & $1.07 \pm 0.20$  & $0.662 \pm 0.015$ & $0.923 \pm 0.019$ & $0.545 \pm 0.011$ \\
$F_{\nu}$ (Jy ms)                      & $3.26 \pm 0.09$   & $3.99 \pm 0.11$  & $3.92 \pm 0.08$   & $6.24 \pm 0.88$  & $2.81 \pm 0.07$ & $3.90 \pm 0.11$  & $3.98 \pm 0.10$   \\
DM $(\rm pc\ cm^{-3})$                 & $541 \pm 10$      & $609 \pm 12$      & $247 \pm 23$      & $299 \pm 42$      & $539 \pm 10$ & $604 \pm 11$      & $398 \pm 28$      \\
eDM $(\rm pc\ cm^{-3})$                & $469 \pm 11$      & $542 \pm 9$      & $151.0 \pm 1.1$   & $243 \pm 41$      & $465 \pm 10$ & $537 \pm 9$      & $323 \pm 25$      \\
$\Delta \nu$ (MHz)                     & $147 \pm 2$   & $400$ & $128 \pm 2$   & $381 \pm 19$      & $160 \pm 2$ & $400$ & $172 \pm 10$      \\
$\log_{10} \Delta t_\mathrm{WS}$ (s)  & $-2.97 \pm 0.01$  & $-3.23 \pm 0.01$ & $-2.64 \pm 0.02$  & $-2.88 \pm 0.12$ & $-3.21 \pm 0.01$ & $-3.24 \pm 0.01$ & $-2.65 \pm 0.02$  \\
$\log_{10} \Delta t_\mathrm{ST}$ (s)  & $-2.07 \pm 0.03$  & $-1.90 \pm 0.02$ & $-2.00 \pm 0.05$  & $-1.78 \pm 0.16$ & $-2.09 \pm 0.03$ & $-1.90 \pm 0.02$ & $-1.99 \pm 0.04$  \\
$\log_{10} E_{\rm iso}$ (erg)          & $39.60 \pm 0.02$  & $39.90 \pm 0.02$ & $39.30 \pm 0.05$  & $39.50 \pm 0.17$ & $39.70 \pm 0.03$ & $39.90 \pm 0.02$ & $39.30 \pm 0.04$  \\
$\log_{10} L_{\rm iso}$ ($\rm erg\,s^{-1}$) & $42.00 \pm 0.02$  & $42.50 \pm 0.03$ & $41.50 \pm 0.06$  & $42.00 \pm 0.20$ & $42.20 \pm 0.03$ & $42.50 \pm 0.03$ & $41.50 \pm 0.06$  \\
$\log_{10} T_{\rm B}$ (K)              & $36.20 \pm 0.04$  & $36.90 \pm 0.03$ & $35.20 \pm 0.08$  & $36.1 \pm 0.50$   & $36.50 \pm 0.03$ & $36.90 \pm 0.03$ & $35.20 \pm 0.08$  \\
\hline \hline
\end{tabular*}
\end{table*}

The parameter $r$ characterizes the curvature of the emission spectrum around the central frequency and is therefore directly linked to the effective emission bandwidth. Values of $r \sim 0$ correspond to relatively flat, broadband spectra, whereas $r < 0$ indicates pronounced narrowband emission. We find that FRBs in the nonrepeater-like cluster predominantly concentrate around $r \sim 0$, consistent with broadband spectral characteristics. In contrast, the repeater-like cluster preferentially occupies the $r < 0$ region, exhibiting significantly narrower spectral features. This behavior aligns with the distribution of the frequency bandwidth $\Delta\nu$, where the nonrepeater-like cluster shows a median bandwidth of $400$ MHz compared to $147$ MHz for the repeater-like cluster.

In terms of temporal properties, FRBs in the repeater-like cluster display systematically longer pulse widths $\Delta t_{\mathrm{WS}}$ than those in the nonrepeater-like cluster. Taken together, the combination of longer durations and narrower bandwidths constitutes a robust observational signature of the repeater-like population. This coherent set of properties naturally explains the pronounced clustering and clear separation of this population in the UMAP embedding space (Figure~\ref{fig:figure1}).

In addition, the nonrepeater-like cluster systematically exhibits higher values of $S_{\nu}$, revealing a clear distinction in the characteristic energy scales of the two populations. To further quantify differences in radiative intensity and emission efficiency, we compare the median logarithmic values of $E_{\mathrm{iso}}$, $L_{\mathrm{iso}}$, and $T_{\mathrm{B}}$.

As listed in Table~\ref{tab:tab1}, the nonrepeater-like cluster displays substantially higher median energies and luminosities, with $E_{\rm iso} = (8.47 \pm 0.42) \times 10^{39}$ erg and $L_{\rm iso} = (3.17 \pm 0.19) \times 10^{42}$ erg s$^{-1}$. In comparison, the repeater-like cluster exhibits lower median values of $E_{\rm iso} = (3.73 \pm 0.19) \times 10^{39}$ erg and $L_{\rm iso} = (1.00 \pm 0.06) \times 10^{42}$ erg s$^{-1}$. Similarly, the brightness temperature $T_{\mathrm{B}}$ is higher for the nonrepeater-like cluster ($T_{\rm B} = (8.39 \pm 0.55) \times 10^{36}$ K) compared to the repeater-like cluster ($T_{\rm B} = (1.46 \pm 0.12) \times 10^{36}$ K). These pronounced differences indicate that nonrepeater-like FRBs typically release more energetic and luminous bursts, whereas repeater-like FRBs tend to operate at comparatively lower energy levels.

Intriguingly, the nonrepeater-like cluster exhibits slightly higher DMs than the repeater-like cluster. This discrepancy becomes significantly more pronounced when considering only confirmed repeaters and the nonrepeating bursts within the nonrepeater-like cluster. A detailed comparison and its possible reasons are elaborated upon in Sec.~\ref{threshold}.

Taken together, the parameter-distribution comparisons and AD test results indicate that the repeater-like cluster corresponds to a distinct phenomenological bursting regime within the FRB population, characterized by narrower spectral bandwidths and lower radiated energies, whereas the nonrepeater-like cluster is dominated by broadband, high-energy bursting modes.

\subsubsection{Parameter comparison of typical and atypical repeaters} \label{sssec:atypical}

During the UMAP embedding of Catalog 2 sample, we find that bursts from repeating sources are separated into two distinct clusters, indicating systematic heterogeneity within the repeater population. It is important to emphasize that this classification distinguishes individual burst events rather than source populations, revealing a statistical separation that can occur within the activity of a single source.

As detailed in Table~\ref{tab:tab1} and Appendix B Figure~\ref{fig:figure10}, we compare the properties of these typical and atypical repeating bursts. The most significant divergence lies in their spectral morphology and energy release. Atypical repeaters exhibit broadband spectra and significantly enhanced energetic parameters ($E_{\mathrm{iso}}$, $L_{\mathrm{iso}}$, and $T_{\mathrm{B}}$), entering a high-energy regime phenomenologically indistinguishable from apparent nonrepeating bursts.

Taken together, the presence of atypical repeating bursts indicates that repeating FRB sources do not operate within a single radiative mode. Moreover, since these atypical bursts are phenomenologically indistinguishable from apparent nonrepeaters, this prompts us to reassess the physical implications of the repeater-like and nonrepeater-like clusters identified earlier. The nonrepeater-like cluster does not necessarily represent a population of distinct physical origin. Instead, this two-cluster structure implies that the observed division might merely reflect varying emission conditions rather than intrinsically different source classes.

We further compare morphologically similar subgroups within the two clusters (parameters detailed in Table~\ref{tab:tab1}): typical repeaters versus narrowband nonrepeaters (assigned to the repeater-like cluster), and atypical repeaters versus broadband nonrepeaters (assigned to the nonrepeater-like cluster). While both nonrepeater sets exhibit slightly shorter durations and higher energies than their repeating counterparts, the most prominent distinction is observed in the extragalactic dispersion measure (eDM). Both broadband and narrowband nonrepeaters present a median eDM $\sim 300\ \mathrm{pc\ cm^{-3}}$ higher than their repeating counterparts, indicating that known repeaters are systematically closer. Combined with the analysis in Section~\ref{miss}, even though these apparent nonrepeaters strongly resemble repeaters morphologically---suggesting they are potential repeaters---the failure to detect their repeating activity is completely consistent with expectations, constrained by their greater distances and insufficient follow-up observations.

\subsubsection{Parameter comparison of all repeaters and broadband nonrepeaters} \label{sssec:allrepeat}

To robustly investigate the intrinsic physical differences between repeating and nonrepeating FRBs, careful sample selection is essential. The currently observed nonrepeater population is unavoidably contaminated by latent repeaters due to limited instrumental sensitivity and observing time \cite{2023PASA...40...57J,2024MNRAS.52711158Y}. Such sample contamination can significantly distort the parameter distributions, thereby hindering the interpretation of their underlying physical nature.

To mitigate this, we focus on the nonrepeater-like cluster identified in this work, which exhibits the most pronounced morphological and physical divergence from the repeater population. We directly compare the broadband nonrepeaters with the entire population of repeating sources, which includes all confirmed repeaters regardless of their cluster assignment. As summarized in Table~\ref{tab:tab1} (with detailed parameter distributions presented in Appendix B Figure~\ref{fig:figure11}), this comparison reveals that broadband nonrepeaters consistently display higher $E_{\rm iso}$, $L_{\rm iso}$, and $T_{\rm B}$, alongside characteristic broadband emission. In contrast, repeaters predominantly occupy a lower-energy regime with narrowband emission. Additionally, broadband nonrepeaters also exhibit significantly higher DMs. The potential underlying causes for these observational discrepancies are discussed in detail in Section~\ref{Unifying}.

\section{Unifying the FRB population through selection effects}\label{Unifying}

\subsection{Detection thresholds and distance-dependent selection} \label{threshold}

In this section, we utilize the systematic divergence between broadband nonrepeaters and the complete repeating population as a primary framework to understand the physical origins of FRBs. As detailed in Section~\ref{sssec:allrepeat}, these two groups demonstrate significant differences across multiple physical properties. Although these discrepancies might superficially imply diverse emission mechanisms \cite{2022ApJ...926..206Z} or even distinct physical progenitors (with nonrepeaters potentially originating from catastrophic events \cite{2013PASJ...65L..12T,2016ApJ...822L...7W,2021MNRAS.501.3184S,2014A&A...562A.137F,2014ApJ...780L..21Z,2019PhR...821....1P}, as opposed to repeaters powered by non-destructive magnetars), recent evidence from burst energy distributions strongly points towards a shared physical origin \cite{2024NatAs...8..337K,2026MNRAS545f1937O,Beniamini2025}. Building upon this unified central engine scenario, we argue that the apparent dichotomy is more naturally explained by strong observational selection effects acting on a unified population. The most critical evidence supporting this interpretation comes from the substantially different DM distributions of the two groups.

\begin{figure}[ht!]
\centering
\includegraphics[angle=0,scale=0.42]{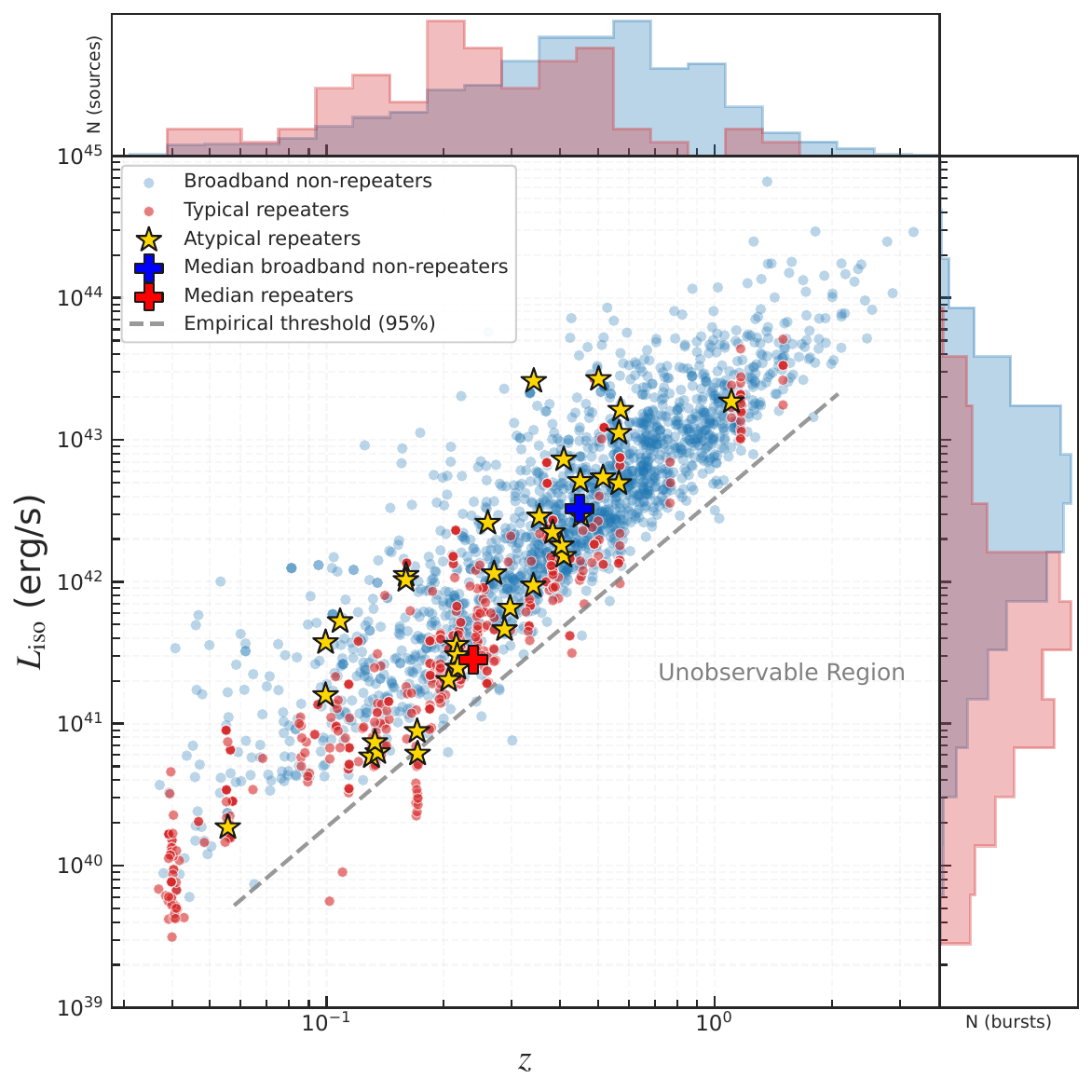}
\caption{Distribution of the FRB sample in the isotropic luminosity versus redshift ($z$--$L_{\rm iso}$) plane. The broadband nonrepeaters, typical repeaters, and atypical repeaters identified in this work are represented by blue dots, red dots, and gold stars, respectively. The marginal histograms on the top and right panels display the normalized distributions of $z$ and $L_{\rm iso}$ for the broadband nonrepeater (blue) and repeater (red) populations, with their respective medians indicated by bold crosses. The dashed curve marks the empirical lower boundary enclosing 95\% of the observed sample. It approximates the effective detection threshold of the CHIME survey.
\label{fig:figure7}}
\end{figure}

Broadband nonrepeaters exhibit a significantly higher median DM ($604 \pm 11~\mathrm{pc\,cm^{-3}}$) compared to repeaters ($398 \pm 28~\mathrm{pc\,cm^{-3}}$), indicating that they are detected at much greater cosmological distances, consistent with previous studies \cite{2023ApJ...947...83C}. In principle, such a disparity in DM could also reflect differences in host galaxy types or local environment. Nevertheless, such factors are not essential to explain this trend. This distance disparity introduces a strong Malmquist bias \cite{1922MeLuF.100....1M}, visible in the $z$--$L_{\rm iso}$ plane (Figure~\ref{fig:figure7}). The normalized marginal histograms and median markers enable a quantitative comparison between broadband nonrepeaters (blue) and repeaters (red). While repeaters are concentrated at lower median redshifts ($z = 0.25 \pm 0.02$) and luminosities ($L_{\mathrm{iso}} = (3.31 \pm 0.41) \times 10^{41}~\mathrm{erg~s^{-1}}$), broadband nonrepeaters extend to higher redshifts ($z = 0.45 \pm 0.01$) and luminosities ($L_{\mathrm{iso}} = (3.28 \pm 0.20) \times 10^{42}~\mathrm{erg~s^{-1}}$).

The dashed curve in Figure~\ref{fig:figure7} delineates the empirical sensitivity threshold of the CHIME telescope, defined such that 95\% of the observed bursts lie above this boundary. This boundary illustrates the effective detection limit, which naturally rises with distance due to the flux-limited nature of the survey. As a consequence, CHIME becomes progressively insensitive to low-luminosity bursts at higher redshifts. For bursts with luminosities comparable to those commonly observed from nearby repeaters, placing them at $z > 0.45$ would cause a substantial fraction to fall below the detection threshold, rendering them effectively unobservable (the unobservable region). As expected from observational selection, the sources detected at higher redshifts are predominantly nonrepeaters. This trend reflects the fact that the high-luminosity nature of these distant events is a prerequisite for their detection. The high-redshift broadband nonrepeaters therefore likely represent the luminous tip of a much larger, unseen population \cite{2019MNRAS.484.5500C,2024NatAs...8..337K,2026MNRAS545f1937O}.

Beyond limiting detectability, distance also severely suppresses the ability to confirm repetition. As discussed by Ref.~\cite{2021A&A...647A..30G}, classification as a repeater requires the detection of at least two bursts from the same source. If the intrinsic FRB luminosity function is steep---such that faint bursts are far more numerous than bright ones---the joint probability of detecting at least two bursts scales roughly as the square of the single-detection probability ($P_{\rm rep} \approx P_{\rm det}^2$). Consequently, this probability declines much more rapidly with distance than that of detecting a single event. Furthermore, when accounting for the stochastic nature of FRB activity (e.g., temporal clustering \cite{2021Natur.598..267L,2022Natur.611E..12X,2023ApJ...955..142Z,2025arXiv250714707Z} or long periods of quiescence \cite{2025MNRAS.540.3709U,2026ApJ...996...47Y}), the effective $P_{\rm rep}$ within a limited survey duration is suppressed to values well below this theoretical approximation. For a nearby source, even faint, frequent bursts are visible, making repetition confirmation relatively easy. In contrast, for distant sources, only the rare, bright tail of the luminosity function is detectable, while the majority of weaker bursts fall below the sensitivity limit. As a result, intrinsically repeating sources at high redshift are statistically predisposed to appear as nonrepeaters.

In summary, high-DM sources are statistically biased toward a nonrepeater classification, even if they arise from intrinsically repeating FRB sources. Together, these distance-dependent selection effects provide a natural explanation for the observed contrasts in DM, luminosity, and repetition behavior.

Crucially, the identification of atypical repeaters (marked by gold stars) offers direct physical evidence bridging the apparent divide between repeating and nonrepeating FRB populations. These events preferentially correspond to the brighter bursts within repeating sources, with this tendency becoming increasingly pronounced at higher redshifts ($z > 0.5$). Several atypical repeaters exhibit isotropic luminosities exceeding $10^{43}$ erg s$^{-1}$, directly overlapping with the luminosity range occupied by broadband nonrepeaters. This overlap demonstrates that repeating FRB sources are intrinsically capable of producing the extreme luminosities previously attributed exclusively to nonrepeating events.

Taken together, the strong distance-dependent selection effects, coupled with instrumental sensitivity limits and the existence of atypical repeaters establish a continuous connection across the $z$--$L_{\rm iso}$ plane. Although the statistical separation between all repeaters and broadband nonrepeaters is robust, it does not require distinct progenitor classes. Instead, the observed differences are consistent with a unified FRB population, in which nearby sources are detected primarily through frequent, low-luminosity bursts and are readily classified as repeaters, while at larger distances only rare, extreme-luminosity bursts exceed the detection threshold, causing intrinsically repeating sources to appear as one-off events. In this framework, broadband nonrepeaters represent the distant, observationally truncated counterparts of luminous repeating sources, rather than a physically distinct population \cite{2019MNRAS.484.5500C,2020MNRAS.494..665L,2023PASA...40...57J}. Future observations with substantially improved sensitivity will be critical for testing this scenario, as they will enable direct access to the faint end of the luminosity function and reveal the currently unseen population of faint, repeating bursts at high redshifts.

\subsection{Missed repetition probability for apparent nonrepeaters} \label{miss}

In addition to missed detections caused by limited instrumental sensitivity, insufficient exposure time and the intrinsic randomness of FRB emissions are key factors that make repeating FRBs difficult to confirm. Here, we conduct a quantitative analysis using the nonrepeaters located within the repeater-like cluster in Figure~\ref{fig:figure1} as an example, estimating the probability of not detecting a second burst, assuming they are intrinsically repeaters. We denote this non-detection probability as $P(\mathrm{miss})$. We adopt the standard Poisson process as our baseline model for this calculation.\footnote{The Weibull distribution is not adopted here to model clustered bursts, as its non-detection probability relies on the discrete durations of individual observing scans, which are unavailable for our CHIME sample. Treating the aggregated cumulative exposure time ($T_{\mathrm{exp}}$) as a single continuous window would introduce systematic biases.}

For an intrinsic repeating source with a burst rate $r$ (in $\mathrm{h}^{-1}$), the probability of not detecting any bursts during a cumulative exposure time $T_{\mathrm{exp}}$ follows Poisson statistics:
\begin{equation}
\begin{aligned}
P(\mathrm{miss} \mid r, T_{\mathrm{exp}}) = e^{-r \cdot T_{\mathrm{exp}}}.
\end{aligned}
\end{equation}

Since the intrinsic burst rate $r$ of each FRB is unknown, we adopt 
the repeating rate prior distribution derived by 
Ref.~\cite{2024MNRAS.52711158Y}, which follows a power-law form:
\begin{equation}
p(r) = \frac{1-q}{r_{\max}^{1-q} - r_{\min}^{1-q}} \cdot r^{-q}, 
\quad r \in [r_{\min}, r_{\max}],
\end{equation}
with the best-fit parameters $q = 2.67$, 
$r_{\min} = 10^{-3.5} \, \mathrm{h}^{-1}$, and 
$r_{\max} = 10^{0.5} \, \mathrm{h}^{-1}$, where the normalization factor ensures $\int_{r_{\min}}^{r_{\max}} p(r)\,\mathrm{d}r = 1$. 
Although this prior was inferred from the full CHIME FRB population, we assume it applies equally to the repeater-like subsample studied here. Marginalizing over this prior yields the expected non-detection probability:
\begin{equation}
P(\mathrm{miss} \mid T_{\mathrm{exp}}) = 
\int_{r_{\min}}^{r_{\max}} e^{-r \cdot T_{\mathrm{exp}}} \cdot p(r) 
\, \mathrm{d}r.
\end{equation}

\begin{figure}[ht!]
\centering
\includegraphics[angle=0,scale=0.6]{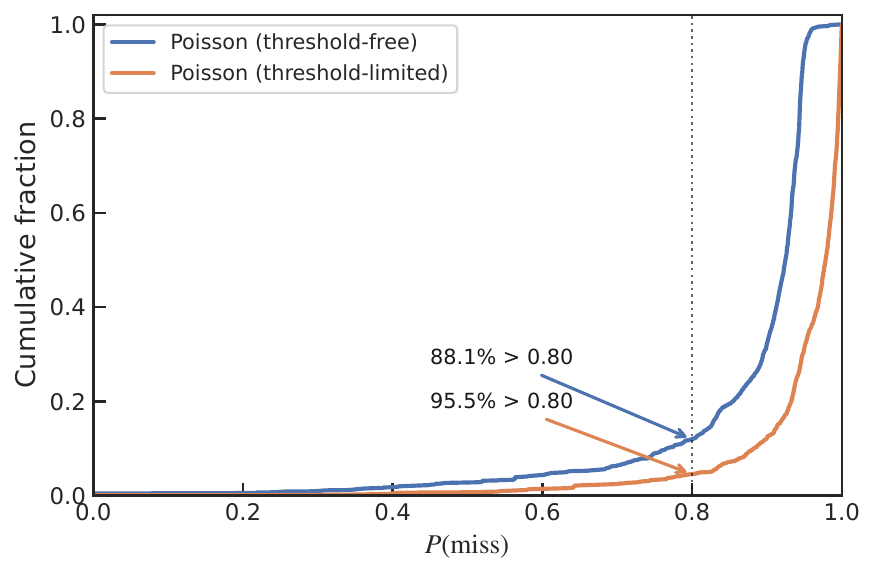}
\caption{Cumulative distribution functions (CDFs) of the non-detection probability, $P(\mathrm{miss})$, for the apparent nonrepeaters distributed within the repeater-like cluster. The calculations are based on the Poisson bursting model. The solid blue and orange curves represent an ideal threshold-free baseline (without selection effects) and a threshold-limited model (incorporating observational selection effects), respectively. The vertical dotted line at $P(\mathrm{miss}) = 0.80$ and the corresponding arrows indicate the fraction of sources with a highly probable missed detection ($P(\mathrm{miss}) > 0.80$).
\label{fig:figure8}}
\end{figure}

Based on the theoretical framework described above, we first calculate the non-detection probability under a threshold-free assumption. This ideal scenario presumes that the detector possesses perfect sensitivity and can seamlessly capture all intrinsic bursts within the exposure window. As shown in Figure~\ref{fig:figure8}, the solid blue line illustrates the cumulative distribution of $P(\mathrm{miss})$ for the FRBs under the Poisson model. The results indicate that even under this idealized framework, the overall sample already exhibits a remarkably high $P(\mathrm{miss})$, with 88.1\% of the sources yielding $P(\mathrm{miss}) > 0.80$. This suggests that the strictly limited cumulative exposure time for individual FRBs currently constitutes a primary observational bottleneck in confirming their repeating nature.

However, this threshold-free assumption is incomplete for realistic 
observations. For FRBs at a given redshift, owing to the intrinsic 
energy distribution of single bursts, the instrument's finite detection 
threshold inevitably truncates a substantial fraction of low-energy 
events. Therefore, CHIME can only detect bursts whose intrinsic energies 
exceed the minimum equivalent energy $E_{\mathrm{th}}(z)$ corresponding 
to the flux threshold $F_{\mathrm{th}}$. Following Ref.~\cite{2025ApJ...990..175L}, the energy--fluence relation is given by:
\begin{equation}
E_{\mathrm{th}}(z) = \frac{4\pi d_L^2(z)}{(1+z)^{2+\alpha}} 
\cdot F_{\mathrm{th}} \cdot \Delta\nu,
\end{equation}
where $d_L(z)$ is the luminosity distance, $\alpha = -1.43$ is the 
spectral index, $\Delta\nu = 1\,\mathrm{GHz}$ is the reference 
bandwidth adopted following Ref.~\cite{2025ApJ...990..175L}, 
and $F_{\mathrm{th}} = 3.5\,\mathrm{Jy\cdot ms}$ is the median 
detection threshold of Catalog 2 \cite{catalog2}. 
We adopt this energy convention to remain consistent with the 
Schechter function parameters used below.

We utilize a Schechter function to model the single-burst energy distribution of FRBs \cite{2025ApJ...990..175L}:
\begin{equation}
P(E) \propto \left(\frac{E}{E_c}\right)^{\gamma} 
\exp\left(-\frac{E}{E_c}\right),
\end{equation}
where the power-law index is $\gamma = -1.49$ and the cutoff energy is $\log_{10}(E_c/\mathrm{erg}) = 41.45$. The integration range is set to 
$E \in [E_{\min},\, E_{\max}]$, where $E_{\min} = 10^{39}\,\mathrm{erg}$ and $E_{\max} = 10^{43}\,\mathrm{erg}$.

Consequently, we define a detection efficiency factor $f(z)$ to quantify the fraction of all intrinsic bursts from a source at a given redshift that can actually be detected by CHIME:
\begin{equation}
f(z) = \frac{\int_{E_{\mathrm{th}}(z)}^{E_{\max}} P(E) \,\mathrm{d}E}{\int_{E_{\min}}^{E_{\max}} P(E) \,\mathrm{d}E},
\end{equation}
where $f(z) = 1$ when $E_{\mathrm{th}}(z) \leq 10^{39}\,\mathrm{erg}$, and $f(z) = 0$ when $E_{\mathrm{th}}(z) \geq 10^{43}\,\mathrm{erg}$.

Substituting this into our model yields an effective exposure time $T_{\mathrm{eff}} = f(z) \cdot T_{\mathrm{exp}}$. Thus, the non-detection probability, accounting for both the limited exposure time and instrumental sensitivity constraints, becomes:
\begin{equation}
P(\mathrm{miss} \mid T_{\mathrm{exp}}, z) = \int_{r_{\min}}^{r_{\max}} P(\mathrm{miss} \mid r, f(z) \cdot T_{\mathrm{exp}}) \cdot p(r) \,\mathrm{d}r.
\end{equation}

As shown in Figure~\ref{fig:figure8}, when the observational sensitivity limit (i.e., the threshold-limited model) is incorporated into the calculations, the cumulative distribution curve of the non-detection probability shifts significantly and systematically to the right (see the solid orange line). At a fiducial non-detection probability of 0.80, our sensitivity-corrected model reveals that 95.5\% of the FRBs under the Poisson assumption have $P(\mathrm{miss}) > 0.80$.

Overall, for the vast majority of FRBs, even if they are intrinsically repeating sources, failing to observe subsequent bursts under current observational conditions is highly probable. This is governed not only by the inherent randomness and temporal clustering of FRB activity but also by the intrinsic distribution of single-burst energies, which inevitably causes the observational threshold $F_{\mathrm{th}}$ to inevitably truncate a substantial number of low-energy bursts. Our analysis quantitatively confirms that severely limited exposure times, coupled with instrumental sensitivity truncation, are the critical systematic factors driving the widespread phenomenon of apparently nonrepeating FRBs.

\section{Conclusion} \label{sec:concl}
In this work, we analyze the multi-dimensional observational properties of FRBs using the largest homogeneous sample of Catalog 2. We adopt an unsupervised machine learning framework that combines UMAP with HDBSCAN clustering to reclassify 4527 burst events. The method remains stable in the large-sample regime, achieving a recall of 0.94 for known repeaters, and enables a systematic comparison of the statistical properties of different FRB clusters. Our main findings can be summarized as follows.
\begin{enumerate}
    \item 
    Based on an unsupervised analysis of Catalog 2, we find that FRBs exhibit a clear and stable structural separation in the embedding space, primarily dividing into two statistically significant clusters. The repeater-like cluster is characterized by narrowband emission and longer pulse widths, encompassing the majority of known repeating bursts, whereas the nonrepeater-like cluster displays broadband, high-energy, and short-timescale properties. This separation is consistently reflected in both the UMAP low-dimensional representation and the observed dynamic spectra, demonstrating the strong physical interpretability of the classification.

    \item 
    Feature-importance analysis shows that the spectral morphology parameter, the spectral running ($r$), consistently serves as the most discriminating metric between repeating and nonrepeating FRBs, reflecting a fundamental difference in their emission bandwidths. This result further indicates that narrowband emission is an intrinsic and fundamental radiative property of the repeating FRB population.
    
    \item 
    Supported by large-sample statistics, we identify a robust subclass of atypical repeaters. Although these bursts originate from known repeating sources, they occupy the same region of the multi-dimensional parameter space as nonrepeating bursts, exhibiting broadband, short-timescale, and high-energy characteristics, while remaining statistically distinct from the typical repeating bursts produced by the same sources. This result may indicate that a single FRB source can exhibit multiple emission modes, thereby blurring the observational boundary between repeating and nonrepeating FRBs.

    \item 
    We find that the observed distinctions between nonrepeaters and repeaters can be naturally explained by instrumental selection effects. The broadband nonrepeaters exhibit systematically higher DMs (typically by $\sim 200\ \mathrm{pc\ cm^{-3}}$) and isotropic luminosities approximately an order of magnitude larger than repeaters. Detected predominantly at higher redshifts ($z > 0.45$), these sources likely represent the high-luminosity tail of the FRB population. In contrast, typical repeaters are preferentially observed at lower redshifts, where fainter, narrowband bursts remain detectable.

\end{enumerate}

In summary, our results support a unified scenario of the FRB population. The phenomenological division between repeaters and nonrepeaters appears to be a consequence of limited observational sensitivity and the stochastic nature of burst emission, rather than a fundamental divide in progenitor types. Crucially, the identification of atypical repeaters serves as a physical continuum between the two classes, indicating that the repeating FRB population itself can exhibit diverse spectral behaviors. Future observations with higher sensitivity and wider bandwidths will be essential for detecting faint emission from distant bursts, thereby mitigating selection effects and providing a more complete census of the FRB population.

\section*{Data Availability}
The original CHIME/FRB Catalog 2 data underlying this work are publicly available at the official CHIME/FRB database. The dataset generated by our unsupervised machine learning framework, including the detailed classification assignments and the corresponding two-dimensional UMAP embedding coordinates ($x$, $y$) for all analyzed FRB events shown in Figure~\ref{fig:figure1}, is publicly available in Ref.~\cite{Sun2026Dataset}.

\section*{Acknowledgments}
We are grateful to Jia-Wei Luo, Long-Xuan Zhang, and Alice P. Curtin for helpful discussions. We acknowledge the support of the National Natural Science Foundation of China (Grants Nos. 12473001, 12533001, and 12575049), the National SKA Program of China (Grant Nos. 2022SKA0110200, 2022SKA0110203), the China Manned Space Program (Grant No. CMS-CSST-2025-A02), and the 111 Project (Grant No. B16009).  Y.L. acknowledges the support of the National Natural Science Foundation of China (No.12473091). F.-W.Z. acknowledges the support from the National Natural Science Foundation of China (No. 12463008) and the Guangxi Natural Science Foundation (No. 2022GXNSFDA035083). Yong-Kun Zhang is supported by the Postdoctoral Fellowship Program and China Postdoctoral Science Foundation under Grant Number BX20250158, and the National Natural Science Foundation of China (Grant No. 12573096).

\section*{Conflict of Interest}
The authors declare that they have no conflict of interest.

\subsection*{Appendix A: Source-level Dimensionality Reduction and Clustering Analysis}
%\label{appendixA}

In this appendix, we perform a supplementary source-level clustering analysis to evaluate the potential statistical differences relative to the burst-level approach. Since bursts from repeating sources tend to exhibit similar narrowband characteristics, the inclusion of multiple bursts from active repeaters naturally introduces a statistical weighting effect. This intra-source correlation may inflate the evaluated recall for repeater identification. To account for this, we construct a source-level dataset by selecting only the first observed sub-burst to represent each distinct FRB source. This curated sample consists of 83 repeaters and 3190 apparent nonrepeaters. We then apply the identical unsupervised machine learning framework used in the main text to this independent-source sample.

\begin{figure}[ht!]
\centering
\includegraphics[angle=0,scale=0.34]{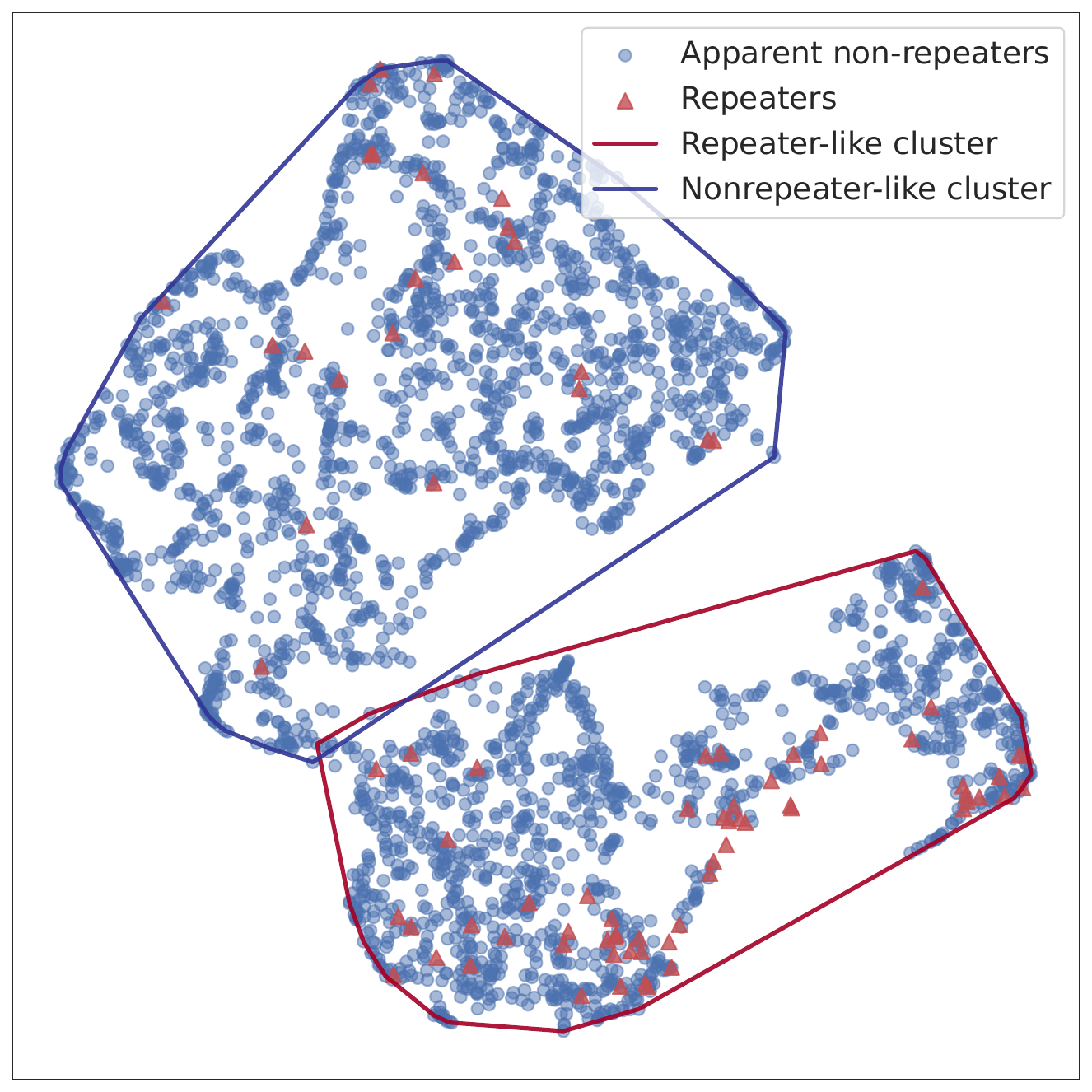}
\caption{Two-dimensional UMAP projection of the source-level FRB sample, with clusters identified by HDBSCAN. Apparent nonrepeaters and repeaters are shown as blue circles and red triangles, respectively. Blue and red contours delineate the HDBSCAN-identified nonrepeater-like and repeater-like clusters.
\label{fig:figure9}}
\end{figure}

Figure~\ref{fig:figure9} presents the resulting UMAP projection and HDBSCAN cluster assignments for the source-level data. In this embedding space, blue circles represent apparent nonrepeaters and red triangles denote the earliest detected event for each repeating source, with all data points uniformly representing the first sub-burst (sub\_num = 0). Maintaining the identical UMAP hyperparameter configuration as in Figure~\ref{fig:figure1}, we find that the dimensionality reduction structure remains highly similar to the burst-level results even after completely removing sample correlation, which demonstrates that our embedding structure is robust. Furthermore, HDBSCAN successfully recovers the two-cluster structure, which consists of a repeater-like cluster (containing the vast majority of repeaters) and a nonrepeater-like cluster (predominantly comprising apparent nonrepeaters). 

Quantitatively, the unsupervised model assigns 60 of the 83 known repeating sources to the repeater-like cluster, with the remaining 23 falling into the nonrepeater-like cluster. According to Equation~\ref{equ:recall}, this yields a source-level recall of 0.72 for repeater identification. Although this value is significantly lower than the recall of 0.94 derived from the burst-level analysis in Figure~\ref{fig:figure1}, this decrease does not indicate a model failure. Instead, it is a highly physically meaningful result. It directly demonstrates that nearly 28\% (23/83) of the repeating sources exhibit morphologies that are virtually indistinguishable from typical nonrepeaters during their initial detection. Furthermore, within the UMAP embedding plane, these 23 repeaters assigned to the nonrepeater-like cluster show no tendency to aggregate locally. Instead, they are uniformly interspersed among the surrounding apparent nonrepeaters, confirming a high degree of morphological similarity.

This significant morphological diversity across classification boundaries implies that we cannot rule out the presence of latent repeaters hidden within the nonrepeater-like cluster. As discussed in Section~\ref{miss}, these sources might simply lack detected repetitions due to greater distances, limited instrumental sensitivity, or insufficient observation time. Therefore, these source-level quantitative results further corroborate our previous interpretation of the two-cluster structure: it does not necessarily represent two fundamentally distinct physical origins, but more likely reflects different emission states of a similar source population operating under varying physical conditions.

\subsection*{Appendix B: Detailed parameter comparisons of FRB subgroups}
\begin{figure*}[t]
\centering
\includegraphics[width=1.0\textwidth]{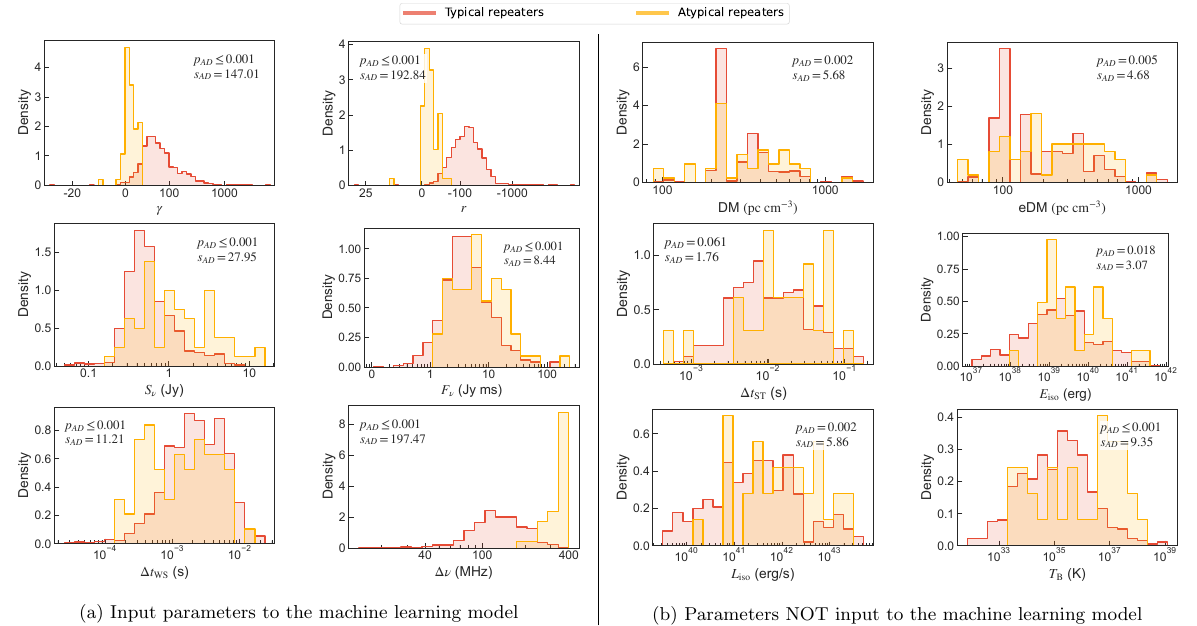}
\vspace{-1ex}
\caption{Parameter distribution comparisons between typical and atypical repeaters. Panel (a) displays the physical parameters utilized as inputs for the UMAP and HDBSCAN classification. Note that although $\Delta \nu$ was not directly input to the model, it is presented here because it is derived from the actual input parameters $\nu_{\rm Low}$ and $\nu_{\rm High}$ ($\Delta \nu = \nu_{\rm High} - \nu_{\rm Low}$). Panel (b) presents the distributions of the parameters not input to the machine learning model. The notations and statistical metrics ($p_{\rm AD}$ and $s_{\rm AD}$) follow Figure~\ref{fig:figure6}.}
\vspace{-2ex}
\label{fig:figure10}
\end{figure*}

%\label{appendixB}
This appendix presents the detailed results of the parameter distributions for the key subgroups determined via clustering analysis. Figure~\ref{fig:figure10} illustrates the detailed parameter distributions for typical and atypical repeating bursts. Consistent with the global separation identified during the UMAP and HDBSCAN clustering process, the most striking visual divergence in Figure~\ref{fig:figure10} occurs within the spectral morphology space. As detailed in Table~\ref{tab:tab1}, typical repeaters are strongly concentrated in the extreme narrowband regime, characterized by median values of $r=-140.1$ and $\gamma=54.1$. Conversely, atypical repeaters exhibit significantly extended distributions toward flatter spectra and lower spectral indices (median $r = -9.6$, $\gamma = 3.9$). These distributions visually confirm that atypical repeating bursts are morphologically and statistically comparable to apparent nonrepeaters.

\begin{figure*}[t]
\centering
\includegraphics[width=0.99\textwidth]{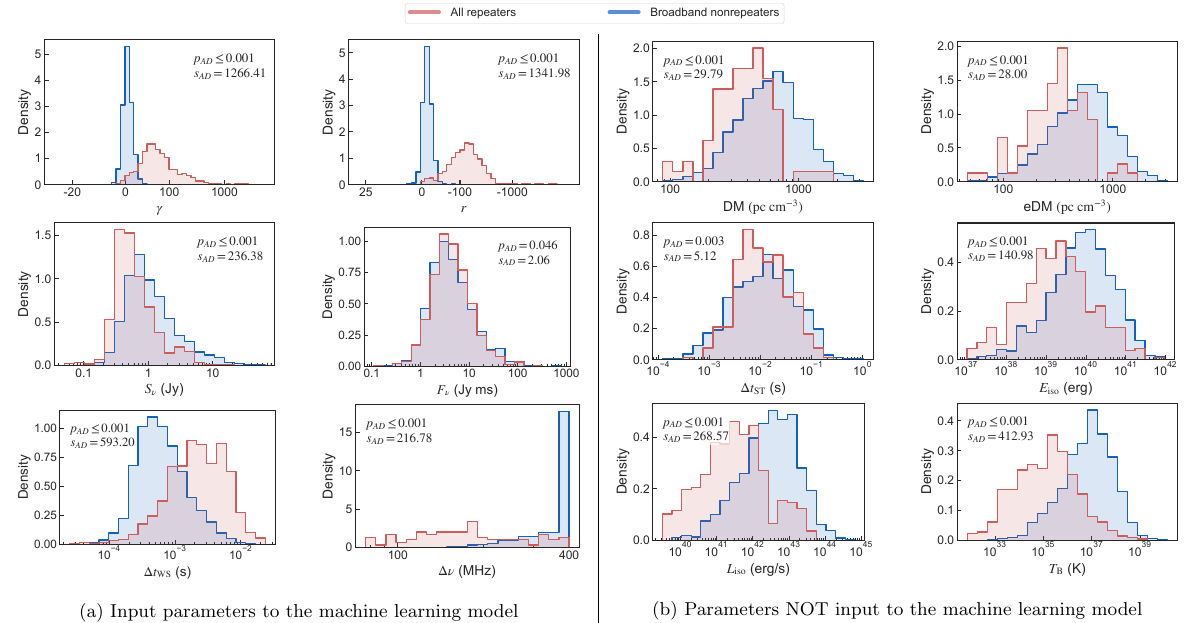}
\vspace{-1ex}
\caption{Parameter distribution comparisons between all repeaters and broadband nonrepeaters. Broadband nonrepeaters are defined as nonrepeaters located within the nonrepeater-like cluster. Panel (a) displays the physical parameters utilized as inputs for the UMAP and HDBSCAN classification. Note that although $\Delta \nu$ was not directly input to the model, it is presented here because it is derived from the actual input parameters $\nu_{\rm Low}$ and $\nu_{\rm High}$ ($\Delta \nu = \nu_{\rm High} - \nu_{\rm Low}$). Panel (b) presents the distributions of the parameters not input to the machine learning model. The notations and statistical metrics ($p_{\rm AD}$ and $s_{\rm AD}$) follow Figure~\ref{fig:figure6}.}
\vspace{-2ex}
\label{fig:figure11}
\end{figure*}

In Figure~\ref{fig:figure11}, we illustrate the parameter distributions of the broadband nonrepeaters compared with the complete population of confirmed repeaters. This direct comparison reveals statistically significant differences ($p_{\mathrm{AD}} \leq 0.001$) across almost all examined physical quantities. However, we note that because multiple bursts originating from the same repeating source share the same physical distance, their derived energetic parameters are not strictly independent samples. This intrinsic correlation affects the strict validity of the AD test to some extent. Therefore, the statistical test results presented here serve primarily as a qualitative reference for evaluating global differences in the overall distributions.

\clearpage
\bibliography{FRB_tSNE}

\end{document}